\documentclass{biophys-new}
\usepackage[utf8]{inputenc}
\usepackage{graphicx}
\usepackage{multirow}
\usepackage{amsmath}
\usepackage{braket}
\usepackage{xcolor}
\makeatletter
\newcommand\setcurrentname[1]{\def\@currentlabelname{#1}}
\makeatother

\usepackage{xr-hyper} 
\usepackage[unicode]{hyperref} 
\hypersetup{
	unicode=true,                                           
	plainpages=false,
	pdffitwindow=true,                                      
	colorlinks=true,                                        
	linkcolor=cyan!70!black,                                        
	citecolor=cyan!70!black,                                         
	filecolor=cyan!70!black,                                        
	urlcolor=cyan!70!black                                          
}

\usepackage{lipsum}

\title{Theoretical study of the influence of the photosynthetic membrane on B800-B850 energy transfer within the peripheral light-harvesting complex LH2}
\runningtitle{Biophysical Journal Template} 

\author[1*]{Chawntell Kulkarni}
\author[2]{Hallmann \'Oskar Gestsson}
\author[3]{Lorenzo Cupellini}
\author[4]{Benedetta Mennucci}
\author[5*]{Alexandra Olaya-Castro}

\runningauthor{Author1 and Author2} 

\affil[1, 2, 5]{Department of Physics and Astronomy, University College London, London WC1E 6BT, United Kingdom}
\affil[3, 4]{Dipartimento di Chimica e Chimica Industriale, Università di Pisa, Via G. Moruzzi 13, 56124 Pisa, Italy}

\corrauthor[*]{chawntell.kulkarni.15@ucl.ac.uk}
\corrauthor[*]{a.olaya@ucl.ac.uk}

\papertype{Article}

\begin{document}

\begin{frontmatter}

\begin{abstract}

Photosynthetic organisms rely on a network of light-harvesting protein-pigment complexes to efficiently absorb sunlight and transfer excitation energy to reaction centre proteins where charge separation takes place. In photosynthetic purple bacteria, such protein-pigment complexes are embedded within the cell membrane, with the lipid composition known to affect the complex clustering, thereby impacting inter-complex excitation energy transfer. However, less is known about the impact of the lipid bilayer on the intra-complex excitation dynamics. Recent experiments have addressed this question by comparing photo-excitation dynamics in detergent-isolated light harvesting complex 2 (LH2) to LH2 complexes individually embedded in membrane discs closely emulating the biological environment. These studies have revealed important differences in spectra and intra-complex energy transfer rates. In this paper we use available quantum chemical and spectroscopy data to develop a complementary theoretical study on the excitonic structure and intra-complex energy transfer kinetics of the LH2 of photosynthetic purple bacteria \textit{Rhodoblastus (Rbl.) acidophilus} (formerly \textit{Rhodopseudomonas acidophila}) in two different conditions: the LH2  in a membrane environment and detergent-isolated LH2.  We find that dark excitonic states, crucial for the B800-B850 energy transfer within the LH2, are more delocalised for the membrane model. By using both non-perturbative and generalised Förster calculations, we show that such increased quantum delocalisation results in a B800 to B850 transfer rate 30\% faster than in the detergent-isolated complex, in agreement with experimental results. We identify the dominating energy transfer pathways in each environment and show how differences in the B800 to B850 transfer rate fundamentally arise from changes in the electronic properties of the LH2 when embedded in the membrane. Furthermore, by accounting for the quasi-static variations of electronic excitation energies in the LH2, we show that the broadening of the distribution of the B800-B850 transfer rates is affected by the lipid composition. We argue that such variation in broadening could be a signature of a speed-accuracy trade-off, commonly seen in biological process.
\end{abstract}

\begin{sigstatement}
Understanding the kinetics of energy transfer within photosynthetic light-harvesting complexes under conditions as close as possible to their biological environments will provide a deeper insight into the biological mechanisms affecting their function. Experiments have shown that for the LH2 complex of photosynthetic purple bacteria, the cell membrane environment can enhance the efficiency of the key energy transfer step within each complex compared to when the photosynthetic complex is isolated via chemical methods. We develop a comprehensive theoretical analysis that rationalises such experimental observations and provide insight into quantum features and microscopic energy transfer pathways that may be enhanced in the membrane environment and which underpin the increased energy transfer rates. 
\end{sigstatement}
\end{frontmatter}

\section*{Introduction}
In purple non-sulphur bacteria, the initial steps of photosynthesis are carried out by a network of protein-pigment complexes which are embedded in the bacterial cell membrane\cite{cogdell2006architecture}. The network is built up of two types of complexes: the light-harvesting complex 2 (LH2) and LH1 which are responsible for the absorption and transfer of incident solar energy and the reaction centre (RC) which accepts excitation energy from the LH1 to facilitate transmembrane charge separation where excitation energy is converted to chemical energy. Since the LH1 surrounds the RC, together they form the core light harvesting complex (LH1-RC). Each LH1-RC is surrounded by several LH2 complexes, forming clusters on the cell membrane \cite{sturgis2008atomic, scheuring2004variable, bahatyrova2004native}.

\begin{figure}[hbt!]
    \centering
    \includegraphics[width=0.9\textwidth]{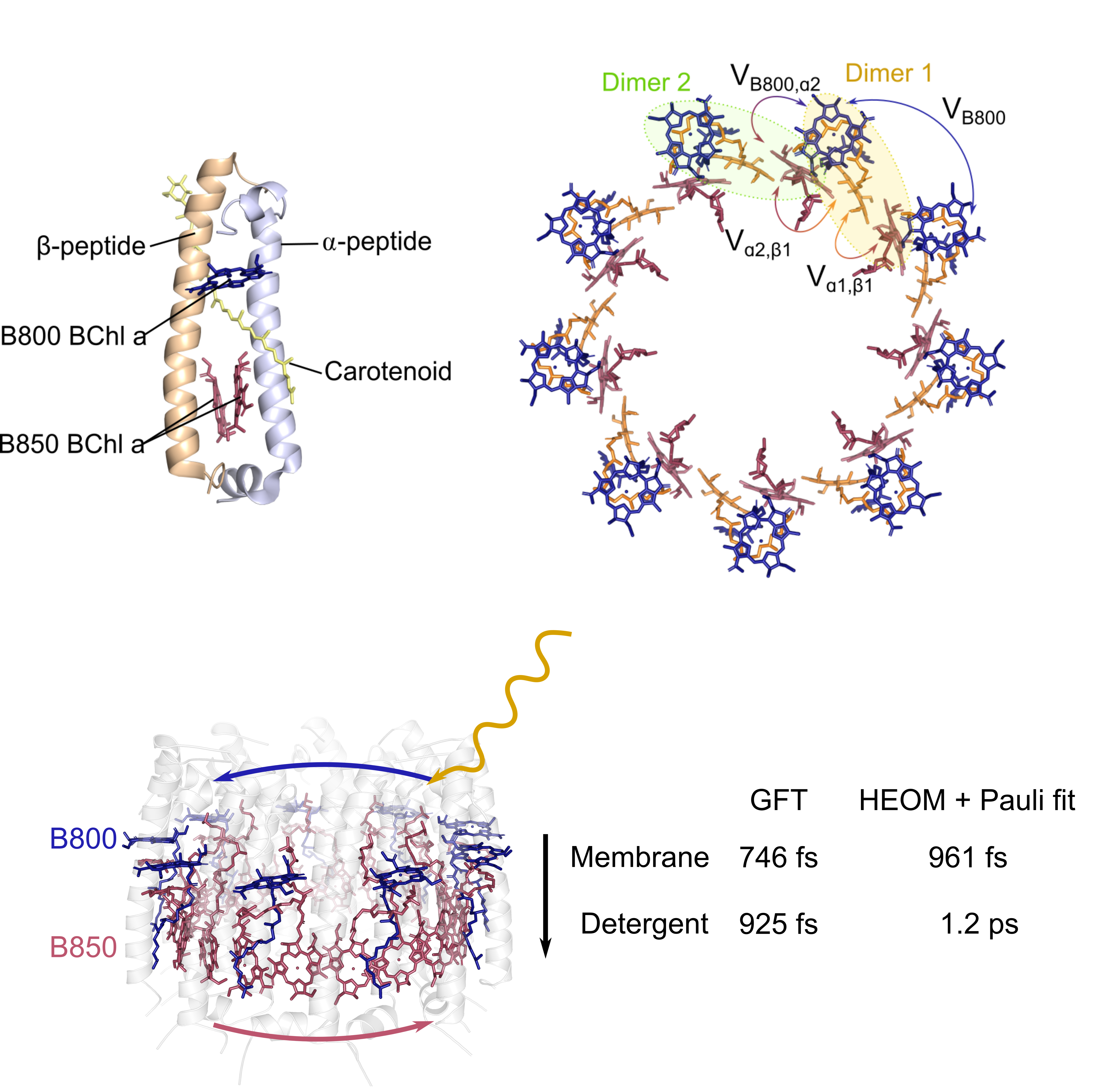}
    \caption{(a) Subunit of the LH2 of \textit{Rbl. acidophilus} (PDB ID: 1NKZ)\cite{papiz2003structure} consisting of an $\alpha\beta$-heterodimer, one B800 and two B850 bacteriochlorophyll a chromophores and one carotenoid. (b) Side view of the LH2 showing the arrangement of the B800 ring (blue) and B850 ring (red) of chromophores. Following the absorption of solar energy by chromophores in the B800 ring, energy is transferred to the B850 ring. Timescales given for interring transfer are computed using generalised F\"orster theory and hierarchical equations of motion. (c) Top view of the chromophores in the LH2, B850$\alpha$ chromophores in red and B850$\beta$ chromophores in orange. Nearest neighbour chromophores couplings are labelled.}
    \label{fig:LH2structure}
\end{figure}

Here we focus on the LH2 from the purple bacteria \textit{Rhodoblastus (Rbl.) acidophilus} which is composed of nine subunits that are arranged in a cyclic C9 symmetry \cite{mcdermott1995crystal, papiz2003structure}. Each subunit consists of one $\alpha\beta$ heterodimer formed from two peptides ($\alpha$ and $\beta$), that bind three bacteriochlorophyll a chromophores (BChl a) and one carotenoid. The Bchl a’s absorb light in the infrared region and are named according to the wavelength of light they approximately absorb at. Each subunit contains one B800 Bchl and two B850 BChl a's labelled $\alpha$ and $\beta$ according to the peptide it is ligated to. Due to the cyclic arrangement of the subunits in the LH2, two concentric rings of chromophores are formed: the B800 ring which lies close to the inner cytoplasmic surface of the membrane and the B850 ring which lies close to the periplasmic surface. The transfer of excitation energy from chromophores in the B800 ring to the B850 ring is a key energy transfer pathway within the LH2 \cite{novoderezhkin2003intra, fidler2013probing, massey2019orientational}. 

Experimental studies focused on understanding the fundamental steps in photosynthetic light harvesting have contributed a vast amount of information on the structure and function of LH2 \cite{cogdell2006architecture, sundstrom1999photosynthetic, georgakopoulou2002absorption, jimenez1996electronic, alden1997calculations, wu1997comparison}. Many of these studies isolate LH2 by solubilising it in detergent, removing it from its native environment in the photosynthetic membrane. The impact of the membrane on the energy transfer dynamics within LH2 remains an open question. Recently, experimental work has found differences in the spectra and energy transfer of detergent isolated LH2 and membrane embedded LH2 \cite{ogren2018impact, freiberg2012comparative, agarwal2002nature, richter2007single}. With the existing comprehensive knowledge on the energy transfer mechanism within detergent isolated LH2, we have a benchmark to perform a systematic study of how energy transfer may be altered when LH2 is embedded in its native membrane environment.

The bacterial photosynthetic membrane is composed primarily of phospholipids with different species of purple non-sulfur bacteria having varying lipid compositions \cite{hunter2009purple, nagatsuma2019phospholipid}. Lipids in the membrane mediate clustering of the LH2 complexes, with different lipid compositions resulting in different clustering tendencies \cite{dewa2013energy, sumino2013influence}. It has been suggested that the difference in organisation of LH complexes can alter the efficiency of energy transfer from initial absorption by an LH2 complex to its arrival at the RC.

Live cells or sections of the native membrane have been studied, but present difficulties due to the complex biological environment \cite{rigaud}. Since whole cells are highly scattering, spectral signals are disturbed when using spectroscopic methods. To circumvent this issue, after isolating the LH2 with detergents, researchers then reconstitute LH2 into an artificial membrane and perform experiments on these samples \cite{pflock2008comparison, pflock2011electronically, richter2007single, dewa2013energy, sumino2013influence, freiberg2012comparative}. Initial studies comparing the spectroscopic properties of detergent solubilised and membrane reconstituted LH2 found little difference between the two, concluding that a single model should be sufficient to describe both scenarios \cite{richter2007single}.

In contrast, experiments comparing LH2 from \textit{Rhodobacter (R.) sphaeroides} solubilised into detergent micelles to LH2 self-assembled into membrane vesicles found differences in the absorption spectra at room temperature \cite{freiberg2012comparative, agarwal2002nature}. In the membrane vesicles, the B850 band of LH2 was broader and red shifted by 1.1 nm and the Stokes shift between the absorption and fluorescence was greater in the membrane. Membrane vesicles typically contain multiple LH2 complexes which, through their intercomplex interactions, can add another environmental contribution to the dynamics of a single LH2 leading to broadening in its spectra. Therefore, to isolate the membrane's effect on the complex, a single LH2 embedded in a membrane is ideal. 

Ogren \textit{et al.} embedded LH2 in a membrane nanodisk which allows a single complex to be separated and probed since each disk holds a single LH2 \cite{ogren2018impact}. The absorption spectra of a single LH2 complex in the membrane nanodisk also exhibited a slightly redshifted B850 absorption peak compared to detergent solubulised LH2 and pump-probe measurements found the B800 to B850 transfer rate in the membrane nanodisk to be ~30\% faster (670 fs) than in detergent (875 fs). 

In this work, we conduct a theoretical study of the impact of the membrane environment on energy transfer within the LH2 of \textit{Rbl. acidophilus}, to determine if the differences in spectra and energy transfer times observed experimentally in \textit{R. sphaeroides} hold across alternate species of purple bacteria and how these differences can be mapped down to microscopic changes in the energy transfer pathways. Atomic level calcualtions for electronic and environmental parameters are currently only available for membrane embedded LH2 from \textit{Rbl. acidophilus} \cite{cupellini2016ab}. However, like \textit{R. sphaeroides}, it contains nine subunits with cyclic C9 symmetry and produces similar linear absorption spectra \cite{kennis1997femtosecond, wu1997comparison} such that its structure is commonly used to model \textit{R. sphaeroides} \cite{tong2020comparison}. Due to these structural and spectral similarities, we aim to see if the changes seen in the spectra and energy transfer times of \textit{R. sphaeroides} can be expected in \textit{Rbl. acidophilus}. We compare two models of LH2, one based on experimental spectra of detergent solubilized LH2 \cite{freiberg2009excitonic} and the other describing LH2 embedded in a 1-palmitoyl-2-oleoyl-glycero-3-phosphocholine (POPC) membrane \cite{cupellini2016ab}. We use two different spectral densities to describe the detergent and membrane environment and calculate energy transfer rates within the LH2 using two different levels of theory: generalised Förster theory (GFT) \cite{scholes2001adapting} a perturbative method and hierarchical equations of motion (HEOM) a numerically exact method. Due to the disordered nature of biological systems, each complex is perturbed differently by its local environment creating slight variations in the electronic properties of each complex. Thus, we use many realisations of the electronic parameters to calculate intercomplex energy transfer rates and exciton properties and analyse the specific form of their statistical distribution to see if they reveal anything about the membrane’s influence on energy transfer dynamics within the LH2. We compare the exciton delocalisation for detergent isolated LH2 and membrane embedded LH2 using the inverse participation ratio as a measure. Using GFT and HEOM, we calculate the B800 to B850 energy transfer rate distribution for both models and consider the B800 and B850 exciton levels that form the dominating energy transfer pathways in each environment. 

\section*{Methods}

\subsection*{Hamiltonian}
To model the LH2 complex, we divide the total system Hamiltonian into the system, the environment and the interaction between the two:
\begin{equation}
    \hat{H} = \hat{H}_{\mathrm{S}}+\hat{H}_{\mathrm{B}}+\hat{H}_{\mathrm{SB}}
    \label{eq:hamiltonian}.
\end{equation}
Here $\hat{H}_{\mathrm{S}}$ represents the electronic degrees of freedom of the $N$ chromophores within the LH2 and is given by a Frenkel exciton Hamiltonian \cite{davydov1964theory}, where each chromophore site is treated as a two level system (we have $\hbar = 1$ throughout),
\begin{equation}
    \hat{H}_{\mathrm{S}} = \sum_{i}^{N} E_{i} \ket{i}\bra{i} + \sum_{i,j<i}^{N}V_{ij} (\ket{i}\bra{j} + \ket{j}\bra{i})
    \label{eq:systemhamiltonian},
\end{equation}
where $\ket{i}$ is an excited state localised on site $i$. $E_{i}=\epsilon_{i} + \lambda_{i}$ is the transition energy from ground to excited state of site $i$ termed the site energy and is the sum of the bare electronic energy in the absence of phonons and the reorganisation energy.  $\lambda_{i} = \pi^{-1}\int_{0}^{\infty} d\omega\ J_{i}(\omega)/\omega$ is the energy the bath must dissipate to relax to the new equilibrium in the excited state $\ket{i}$ which can be obtained by integrating over the spectral density $J_{i}(\omega)$. The microscopic origin of $\lambda_{i}$ is due to the excited state potential energy surface being displaced relative to the ground state \cite{valkunasmolec}. $V_{ij}$ is the electronic coupling between the $Q_{\mathrm{y}}$ transition dipole moments at sites $i$ and $j$. We denote $\ket{\alpha}$, the eigenstates of $\hat{H}_s$ with energy $E_\alpha$, i.e. $\hat{H}_s=e_{\alpha}\ket{\alpha}$, which are collective electronic states, or excitons, delocalised across all chromophores, i.e. $\ket{\alpha}=\sum_i C_i^\alpha \ket{i}$.

Site energies and nearest neighbour electronic couplings for the membrane and detergent Hamiltonian's are given in Table \ref{tab:1}. For the detergent Hamiltonian, interchromophore electronic couplings are calculated using the dipole-dipole approximation,
\begin{equation}
    V_{ij}^{\mathrm{dipole}} = C \frac{\hat{\mathbf{d}}_{i}\cdot\hat{\mathbf{d}}_{j} - 3(\hat{\mathbf{r}}_{ij}\cdot\hat{\mathbf{d}}_{i}) (\hat{\mathbf{r}}_{ij}\cdot\hat{\mathbf{d}}_{j})} {|r_{ij}|^3}
    \label{eq:dipolecouplings},
\end{equation}
where $C$ is a constant accounting for the dipole strength, $\hat{\mathbf{d}}_{i}$ is the transition dipole unit vector at site $i$, $\hat{\mathbf{r}}_{ij}$ is the unit vector pointing from the position of site $i$ to site $j$ and $r_{ij}$ is the distance between sites $i$ and $j$. The site coordinates and transition dipole moments are taken from the crystal structure of LH2 from \textit{Rbl. acidophilus} \cite{papiz2003structure} and $C$ is taken to be 230,000 \r{A}cm$^{-1}$ for the B800 sites and 348,000 \r{A}cm$^{-1}$ for the B850 sites, chosen to reproduce energies of the excitonic states. Additionally, these values of C produce couplings that agree with more sophisticated transition density cube methods used to determine electronic couplings in the LH2 \cite{krueger1998calculation, tretiak2000bacteriochlorophyll}. For nearest neighbour electronic couplings in the B850 ring, the dipole-dipole approximation no longer holds due to the proximity of the chromophores, hence couplings were taken from literature where they are fitted to reproduce experimental spectra \cite{freiberg2009excitonic}. The electronic parameters for the membrane Hamiltonian were calculated using quantum chemical methods that account for the mutual polarisation between the lipid-protein environment and the chromophores \cite{cupellini2016ab, curutchet2011photosynthetic}. Site energies and couplings are averaged over a trajectory of the LH2 in a lipid environment using molecular dynamics simulations. The site energies and nearest neighbour couplings of the B800 and B850 chromophores are taken from \cite{cupellini2016ab} and are given in Table \ref{tab:1}.

\begin{table}[htbp]
    \small
    \centering
    \begin{tabular}{llccc}
    \hline
            &       & Membrane POPC \cite{cupellini2016ab} & Membrane DOPC \cite{ramos2019molecular} & Detergent \cite{freiberg2009excitonic}\\
    \hline
        Site energy     & B800                  & 13021     & 13783     & 12540 \\ 
                        & B850$\alpha$          & 12799     & 13527     & 12390 \\ 
                        & B850$\beta$           & 12806     & 13556     & 12390 \\
    \hline
        B800 couplings  & $V_{\mathrm{B}800}$   & -33       & -34       & -19   \\

        B850 couplings  & $V_{\alpha1\beta1}$   & 339       & 298       & 315   \\
                        & $V_{\alpha2\beta1}$   & 317       & 266       & 245   \\
        B800 to B850 couplings  & $V_{\mathrm{B}800, \alpha2}$   & 42 & 38 & 32 \\
    \hline
        Reorganisation energy   & $\lambda_{\mathrm{B800}}$  & 40    & 40    & 35    \\
                                & $\lambda_{\mathrm{B850}}$  & 140   & 140   & 160   \\
        Cut-off frequency       & $\Omega_{\mathrm{B800}}$   & 100   & 100   & 35    \\
                                & $\Omega_{\mathrm{B850}}$   & 100   & 100   & 53    \\
    \hline
        Static disorder & $\sigma_{\mathrm{B800}}$  & 40    & 40     & 50    \\
                        & $\sigma_{\mathrm{B850}}$  & 270   & 270    & 220   \\
    \hline
    \end{tabular}
\caption{Site energies, nearest neighbour electronic couplings and environmental parameters of the chromophore sites in the LH2 from \textit{Rbl. acidophilus} for the membrane and detergent models given in units of cm$^{-1}$. Membrane POPC parameters were taken from \cite{cupellini2016ab}, DOPC membrane electronic parameters were taken from \cite{ramos2019molecular} and detergent parameters for the B850 ring were taken from \cite{freiberg2009excitonic}. Interchromophore electronic couplings are illustrated in Figure \ref{fig:LH2structure}b.}
\label{tab:1}
\end{table}

The environment, $H_{\mathrm{B}}$, corresponds to the intermolecular vibrations of the chromophores along with the motion of the proteins and is modelled as a bath of quantised harmonic oscillators (vibrational modes),
\begin{equation}
    \hat{H}_{\mathrm{B}}=\sum_{i,k}\omega_{i,k}\left(\hat{b}_{i,k}^{\dagger}\hat{b}_{i,k} + \frac{1}{2}\right)
    \label{eq:bathhamiltoniandiscrete},
\end{equation}
where $b_{i,k}^{\dagger}$ and $b_{i,k}$ are bosonic creation and annihilation operators of frequency modes $\omega_{i,k}$ satisfying commutation relations $[b_{i,k},b_{j,k^\prime}^{\dagger}] = \delta_{i,j}\delta_{k,k^\prime}$ \cite{may2008charge}.
Each site is linearly coupled to an environment displacement mode such that the system-environment interaction is of the form
\begin{equation}
    \hat{H}_{\mathrm{SB}} = \sum_{i,k}g_{i,k}(\hat{b}_{i,k}+\hat{b}_{i,k}^{\dagger})\ket{i}\bra{i} = \sum_{i}\hat{B}_{i}\ket{i}\bra{i}
    \label{eq:interactionhamiltonian},
\end{equation}
where $g_{i,k}$ is the interaction strength.

Influence of the environment on the system dynamics may be described fully by the system-bath correlation function
\begin{equation}
    C_i(t) = \langle \hat{B}_{i}(t)\hat{B}_{j}(0)\rangle_{B} = \frac{1}{\pi}\int_0^\infty\text{d}\omega J_i(\omega)\left(\coth\left(\frac{\beta\omega}{2}\right)\cos(\omega t) - i\sin(\omega t)\right),
    \label{eq:correlationfunction}
\end{equation}
where $\beta = \frac{1}{k_{\mathrm{B}}T}$.

Within each band of LH2, we assume that local electronic-vibrational interactions are identical such that all sites are characterised by the same spectral density which takes the Drude-Lorentz form,
\begin{equation}
    J_{i}(\omega) = 2\lambda_{i}\gamma_{i} \frac{\omega}{\omega^{2} + \gamma_{i}^{2}}
    \label{eq:spectraldensity},
\end{equation}
where $\gamma_{i}$ is the cutoff frequency corresponding to the bath relaxation rate. For a Drude-Lorentz spectral density, the bath correlation function may be expressed as an exponential series \cite{valkunasmolec}
\begin{equation}
    C_i(t) = \sum_k c_{k,i}e^{-\nu_{k,i}t},
    \label{eq:correlationfnexp}
\end{equation}
where the coefficients and rates that enter the expansion are obtained using the Matsubara expansion method, $c_{0,i} = \lambda_i\gamma_i\left(\cot\left(\frac{\beta\gamma_i}{2}\right) - i\right)$, $\nu_{0,i} = \gamma_i$, $c_{k,i} = \frac{4\lambda_i\gamma_i}{\beta}\frac{\nu_k}{\nu_k^2 - \gamma_i^2}$ and $\nu_{k,i} = \nu_k$,
where $\nu_k = \frac{2\pi k}{\beta}$ are the Matsubara frequencies with $k = 1, 2, 3\ldots$. The environmental parameters introduced here, $\lambda_i$ and $\gamma_i$ are given for membrane embedded and detergent isolated LH2 in Table \ref{tab:1}.

\subsection*{Static disorder}
In the previous section, fixed electronic parameters were given for the chromophore sites in the LH2. However, owing to the dynamic nature of the biological environment, slow conformational motions of the proteins lead to  random shifts in the electronic parameters of the chromophores \cite{cogdell2006architecture}. Stochastic fluctuations in the local environment of the chromophores create shifts in their site energies while changes in the orientation and position of the chromophore transition dipole moments which alter interchromophore couplings \cite{jang2001characterization}. Since these changes are slow compared to energy transfer timescales, they can be accounted for by taking an ensemble average over many realisations of the electronic parameters. 

Single molecule spectroscopy has shown that static disorder is largely diagonal \cite{hofmann2004energetic}. Therefore, we account for static disorder by adding an offset $\delta_{i}^{r} \in \{\delta_{i}\}_{r}$ to the site energies of the system Hamiltonian in the chromophore site basis
\begin{equation}
    \hat{H}_{\text{S}}^{r} = \sum_{i}^{N} (E_{i} + \delta_{i}^{r}) \ket{i}\bra{i} + \sum_{i,j<i}^{N}V_{ij} (\ket{i}\bra{j} + \ket{j}\bra{i}),
    \label{eq:disorderhamiltonian}
\end{equation}
where r labels a particular realisation.
Each $\delta_{i}^{r}$ is randomly sampled from a Gaussian distribution centred at zero, whose standard deviation, $\sigma$, corresponds to the level of static disorder. Hence, excitonic energies and exciton delocalisation are different for each realisation.  Calculations of observables are averaged over many realisations of static disorder in order to account for its effects on the system. Static disorder for the B800 sites and B850 sites in detergent and membrane are given in Table \ref{tab:1}.

\subsection*{l1 Norm of Coherence}
Due to strong interchromophore electronic couplings in the B850 ring, an excitation in the ring manifests as a delocalised exciton state spread across multiple chromophore sites. In order to quantify the delocalisation of an exciton state $\ket{\alpha}$, we will use two measures: the l1 norm of coherence \cite{baumgratz2014quantifying} and, the more known, participation ratio. This will allow us to analyse if different quantifiers of exciton delocalisation lead to the same conclusions. 

The l1 norm of coherence denoted as $C_{l1}$ \cite{baumgratz2014quantifying} is a measure of coherence based on distance measures and represents the distance of the density matrix associated to $\bra{\alpha}$ i.e $\hat{\rho}^{\alpha}=\ket{\alpha}\bra{\alpha}$ to the set of incoherent quantum states in the reference basis $\set{\ket{i}}$. $C_{l1}(\hat{\rho}^{\alpha})$ is then given by 
\begin{equation}
C_{l1}(\hat{\rho}^{\alpha})=\sum_{{i,j\neq i}} |\hat{\rho}_{i,j}^{\alpha}|= \sum_{{i,j\neq i}} |C_{i}^{\alpha}(C_{j}^{\alpha})^*|
\label{eq:delocalisation},
\end{equation}
where $C_{i}^{\alpha}=\bra{i}\alpha\rangle$ is the amplitude of the excited state of chromophore $i$ in the exciton eigenstate $\ket{\alpha}$. Under incoherent processes, $C_{l1}$ does not increase and therefore it provides an appropriate quantifier of coherence \cite{streltsov2017colloquium}. 

A more common measure of exciton delocalisation is the inverse participation ratio (IPR) which is given by,
\begin{equation}
    \mathrm{IPR}_{\alpha} = \frac{1}{\sum_{i}^{N}|C_{i}^{\alpha}|^{4}},
\end{equation}
where $C_{i}^{\alpha}$ is as defined above. The IPR represents how many chromophores an exciton state $\ket{\alpha}$ is extended over. For example, for a localised exciton IPR = 1 while for a completely delocalised exciton IPR = $N$, where $N$ is the number of chromophores in the ring.

\subsection*{Hierarchical equations of motion}\setcurrentname{hierarchical equations of motion}\label{Sec:HEOMTheory}

In order to quantify energy transfer rates within the LH2, we apply the hierarchical equations of motion (HEOM) \cite{Tanimura1989Jan, Tanimura2020Jul} to compute the quantum dynamics for the the full 27 site model of LH2 that includes both the B800 and B850 and interactions among them in order to predict linear spectra and estimate transfer rates. The HEOM can yield exact quantitative results for the electronic dynamics provided that system-environment correlation functions are represented by an exponential series expansion as in Eq. \eqref{eq:correlationfunction}.

The HEOM is of the form
\begin{equation}\label{Eq:HEOM}
    \dot{\hat{\rho}}_{\textbf{n}} = \left(\mathcal{L} - \Xi - \sum_{k,i} n_{k,i}\nu_{k,i}\right)\hat{\rho}_{\textbf{n}} - i\sum_{k,i}\left(\mathcal{L}_{k,i}^-\hat{\rho}_{\textbf{n}_{k,i}^-} + \mathcal{L}_{k,i}^+\hat{\rho}_{\textbf{n}_{k,i}^+}\right),
\end{equation}
where $\textbf{n}$ is a multi-index consisting of discrete integers $n_{k,i}$. An auxiliary density operator (ADO) $\hat{\rho}_{\textbf{n}}$ is said to belong the $n$-th tier of the hierarchy if $\sum_{k,i} n_{k,i} = n$. The reduced density matrix of the system is identified as $\rho_{\textbf{0}}$. The hierarchy in Eq.\ \eqref{Eq:HEOM} is formalized in terms of super-operators such that for an arbitrary system operator $\hat{A}$ we may write $\hat{A}^\times$ and $\hat{A}^\circ$ which denote super-operators whose action onto a system space operator $\hat{B}$ is given by $\hat{A}^\times\hat{B} = [\hat{A}, \hat{B}]$ and $\hat{A}^\circ\hat{B} = \{\hat{A}, \hat{B}\}$. We have
\begin{align}
    &\mathcal{L} = -i\hat{H}_{\text{S}}^\times, \\
    &\mathcal{L}_{k,i}^- = \text{Re}(c_{k,i})\hat{n}_i^\times + i\text{Im}(c_{k,i})\hat{n}_i^\circ, \\
    &\mathcal{L}_{k,i}^+ = \hat{n}_i^\times.
\end{align}

We truncate the hierarchy by setting all ADOs beyond a pre-set hierarchy tier to zero. The truncation tier $L$ is simultaneously set to be large enough such that numerical results have converged, and small enough so that the simulation will run in a reasonable amount of time. The Matsubara series is truncated as well by approximating $e^{-\nu_k t}\approx\frac{1}{\nu_k}\delta(t)$ for all $k\geq M$, where $M$ is another pre-set threshold chosen similarly to $L$. These approximated terms for the series expansion are then described by the terminator term $\Xi = \sum_{m}\left(\frac{2\lambda_m}{\beta\gamma_m}\left(1 - \frac{\beta\gamma_m}{2}\cot\left(\frac{\beta\gamma_m}{2}\right)\right) - \sum_{k=1}^M\frac{c_{k,m}}{\nu_k}\right)\hat{n}_m^\times\hat{n}_m^\times$ \cite{Ishizaki2005Dec}. We furthermore improve convergence of the HEOM results by applying the scaling procedure developed by Shi and co-workers \cite{Shi2009Feb}.

\subsubsection*{Exact ring population dynamics and its fit to a Pauli master equation}

In order to estimate B800 to B850 energy transfer rates based on the HEOM dynamics, we take our initial state to be the Boltzmann distribution for the B800 eigenstates, i.e. $\hat{\rho}(0) = \frac{e^{-\beta\hat{H}_{\text{B800}}}}{\text{Tr}(e^{-\beta\hat{H}_{\text{B800}}})}$, which is then propagated in time as per the HEOM in Eq.\ \eqref{Eq:HEOM}. We define the total B800 population dynamics as $P_{B800}(t)=\sum_{\alpha \in B800} \langle \alpha |\hat{\rho}(t)|\alpha\rangle$ with $|\alpha\rangle$ the exciton eigenstates of  $\hat H_{B800}$, and similarly for the total B850 population dynamics,  $P_{B850}(t)$. To estimate the transfer rates from B800 to B850, once a steady state is reached, we fit  $P_{B800}$ and  $P_{B850}$ to a Pauli master equation of the form
\begin{equation}
    \partial_t\begin{bmatrix}
        P_{\text{B800}} \\ P_{\text{B850}}
    \end{bmatrix} = 
    \begin{bmatrix}
        -k_{\text{down}} & k_{\text{up}} \\ k_{\text{up}} & -k_{\text{down}}
    \end{bmatrix}
    \begin{bmatrix}
        P_{\text{B800}} \\ P_{\text{B850}}
    \end{bmatrix},
\end{equation}
where $k_{\text{up}}$ and $k_{\text{down}}$ are uphill and downhill decay rates corresponding to the B850$\rightarrow$ B800 and B800$\rightarrow$ B850 transfer process, respectively. We can solve for $P_{\text{B800}}$ by using the fact that $P_{\text{B800}}(t) + P_{\text{B850}}(t) = 1$ such that the B800 population dynamics is of the form
\begin{equation}
    P_{\text{B800}}(t) = \frac{k_{\text{up}} + k_{\text{down}}e^{-(k_{\text{up}} + k_{\text{down}})t}}{k_{\text{up}} + k_{\text{down}}},
    \label{eq:pauli}
\end{equation} 
where the $k_{\text{down}}$ and $k_{\text{up}}$ are numerically determined from a fit to HEOM-simulated population dynamics. This procedure allows estimation of rates that are qualitatively comparable to GFT rates but we do not expect a full quantitative agreement as we are effectively mapping the kinetics of transfer to a two state system, whereas GFT rates consider a multiple parallel processes of exciton to exciton transfer. We will indeed show the qualitative agreement between HEOM and GFT rates and therefore find that the results from the exact treatment support the insight gained from GFT.

\subsubsection*{Linear spectra}
Linear absorption spectra are computed using
\begin{equation}\label{eq:absorption_spectra_expression}
    \alpha_{\mathrm{A}}(\omega) = \text{Re}\left[\sum_{p=x,y,z}\int_0^\infty \text{d}t \braket{\hat{\mu}_p(t)\hat{\mu}_p(0)}_{\rho_0}e^{i\omega t}\right],
\end{equation}
where the initial state of the system is the ground state $\rho_0 = \ket{0}\bra{0}$ and $\hat{\mu}_p(t)$ is the Heisenberg picture dipole operator corresponding to the $p$-direction. The dipole operators are of the form
\begin{equation}
    \hat{\mu}_p = \sum_i d_{i,p}\ket{i}\bra{0} + \text{h.c.},
\end{equation}
where $d_{i,p}$ is the $p$ component of the the dipole at site $i$. 

Linear fluorescence spectra are computed using,
\begin{equation}\label{eq:fluorescence_spectra_expression}
    I_{\mathrm{D}}(\omega) = \text{Re}\left[\sum_{p=x,y,z}\int_0^\infty \text{d}t \braket{\hat{\mu}_p(t)\hat{\mu}_p(0)}_{\rho_{\text{th}}}e^{i\omega t}\right],
\end{equation}
where the initial state of the system is the thermal steady state of the system. We determine $\rho_{\text{th}}$ iteratively via the biconjugate gradient stabilized method \cite{vanderVorst1992Mar} with an initial guess given by the Boltzmann distribution $\rho(0)=\frac{e^{-\beta H_{B800}}} {\text{Tr}(e^{-\beta H_{B800}})}\oplus\mathbb{I}_{B850}$, where $\mathbb{I}_{B850}$ is the identity for the single excitation subspace of the B850 ring.

\subsection*{Generalised F\"orster theory}\setcurrentname{generalised F\"orster theory}\label{Sec:GFTTheory}
In addition to HEOM, we use GFT to calculate the B800 to B850 energy transfer rate. By doing so, we can confirm that our results hold qualitatively at different levels of theory and are not dependent on the approximations made in GFT. Additionally, GFT is a less computationally expensive method that allows the computation of more realisations of static disorder within a reasonable time frame.

GFT describes exciton energy transfer from a donor aggregate to an acceptor aggregate that is weakly coupled to one another \cite{mukai1999theory, scholes2000mechanism, renger2009theory}. It is assumed that, within each aggregate, electronic couplings are strong such that an excitation forms a delocalised exciton state. In the LH2, the donor and acceptor aggregates correspond to the B800 and B850 rings. Strong interchromophore couplings in each ring allow for an excitation to be delocalised across the ring instead of being confined to a single chromophore site. 

To model B800 to B850 energy transfer, it is assumed that following an electronic transition in the B800 ring, thermal relaxation occurs on a shorter timescale than energy transfer, such that transfer to B850 occurs from a thermally populated B800 state. Thus, the B800 to B850 energy transfer rate is given by \cite{mukai1999theory}:
\begin{equation}
    K_\mathrm{GFT} = \sum_{\alpha,\beta} P_{\alpha}k_{\alpha \beta}
    \label{eq:gft},
\end{equation}
where $\alpha$ labels a donor exciton, $\beta$ labels an acceptor exciton, $P_{\alpha}$ is the thermal population of the donor state and k is the exciton transfer rate from $\alpha$ to $\beta$, which is given by the product of the square magnitude of the exciton coupling and the exciton spectral overlap,
\begin{equation}
    k_{\alpha \beta} = |V_{\alpha \beta}|^2 O_{\alpha\beta}
    \label{eq:forstertheory}.
\end{equation}

V$_{\alpha\beta}$ is given by \cite{renger2009theory},
\begin{equation}
    V_{\alpha\beta} = \sum_{i\in D, j\in A} C_{i}^{\alpha} C_{j}^{\beta*}V_{ij}
\label{eq:excitoncoupling},
\end{equation}
where $C_{i}^{\alpha} = \braket{i|\alpha}$ is the amplitude coefficient of site i in the donor exciton eigenstate.

$O_{\alpha\beta}$ is the spectral overlap between the donor fluorescence line shape $\tilde{D}_{\alpha}(\omega)$ and acceptor absorption line shape $D_{\beta}(\omega)$ given by,
\begin{equation}
    O_{\alpha\beta}(\omega) = \frac{1}{2\pi} \int_{-\infty}^{\infty} \ \mathrm{d}\omega \tilde{D}_{\alpha}(\omega) D_{\beta}(\omega)
    \label{eq:spectraloverlap}.
\end{equation}
The form of the lineshape functions may be obtained using pertubative theories as given in the following section.

\subsubsection*{Lineshape theory}
To determine the lineshapes we follow the method outlined by Renger \cite{renger2002relation} where the second order cumulant expansion is used to derive an equation of motion of the reduced system density matrix. This yields lineshape functions of the form

\begin{equation}
    \tilde{D}_{\alpha}(\omega) = 2\mathrm{Re}\int_{0}^{\infty} \mathrm{d}t\ e^{i\omega t} e^{-i(\omega_{\alpha} - \lambda_{\alpha\alpha,\alpha\alpha})t - g_{\alpha\alpha,\alpha\alpha}^{*}(t) - t/\tau_{\alpha}},
    \label{eq:fluorline}
\end{equation}
\begin{equation}
    D_{\beta}(\omega) = 2\mathrm{Re}\int_{0}^{\infty} \mathrm{d}t\ e^{i\omega t} e^{-i(\omega_{\beta} - \lambda_{\beta\beta,\beta\beta})t - g_{\beta\beta,\beta\beta}(t) - t/\tau_{\beta}},
    \label{eq:absline}
\end{equation}
for which $O_{\alpha\beta}$ may be written as:
\begin{equation}
\begin{aligned}
O_{\alpha\beta}(\omega) = 2\mathrm{Re}\int_{0}^{\infty} \mathrm{d}t & e^{i\omega_{\alpha\beta}t} e^{-i(\lambda_{\alpha\alpha,\alpha\alpha}+\lambda_{\beta\beta,\beta\beta})t} \\
&\times e^{-(g_{\alpha\alpha,\alpha\alpha}(t)+g_{\beta\beta,\beta\beta}(t))} e^{-(1/\tau_{\alpha}+1/\tau_{\beta})t}
\label{eq:spectraloverlapfull},
\end{aligned}
\end{equation}
where $\omega_{\alpha}$ is the energy of exciton $\alpha$, $\lambda_{\alpha\beta,\gamma\delta} = \sum_{i}(C_{i}^{\alpha})^{*}C_{i}^{\beta}(C_{i}^{\gamma})^{*}C_{i}^{\delta} \lambda_{i}$ is the exciton reorganisation energy, $g_{\alpha\beta,\gamma\delta}(t) = \sum_{i}(C_{i}^{\alpha})^{*}C_{i}^{\beta}(C_{i}^{\gamma})^{*}C_{i}^{\delta} g_{i}(t)$ is the exciton line broadening function and $\tau_{\alpha}$ is the lifetime of exciton $\alpha$. The exciton lifetimes are approximated using modified Redfield theory as outlined in the supporting material. $g_{i}(t)$ is the site line broadening function which, for the bath correlation function we consider (Eq. (\ref{eq:correlationfnexp})), may be written as
\begin{equation}
    g_{i}(t) = \frac{c_{0,i}}{\gamma_i^{2}}(e^{-\gamma_it} + \gamma_it-1) + \sum_{k=1}\frac{c_{k,i}}{\nu_{k}^{2}}(e^{-\nu_{k}t} + \nu_{k}t-1),
    \label{eq:sitelbf}
\end{equation}
The Matsubara summation terms labelled by $k$ are low temperature corrections to the exponential expansion of the bath correlation function. Since we are interested in the function of LH2 in a physiological environment, our calculations are at 300K where the Matsubara terms are less important, hence we we truncate the summation at $k=1$, as the correlation function $C_{i}$ does not change when including higher order terms.

Aside from computing energy transfer rates, the lineshapes in eqs. (\ref{eq:absline}) and (\ref{eq:fluorline}) are also used to compute linear spectra. The linear absorption and fluorescence spectra of the respective B800 and B850 rings can be obtained using their relationship to $D_{\beta}(\omega)$ and $\tilde{D}_{\alpha}(\omega)$ \cite{renger2009theory},
\begin{equation}
    \alpha_{A}(\omega) \propto \sum_{\beta}|\vec{\mu}_{\beta}|^2 D_{\beta}(\omega),
\end{equation}
\begin{equation}
    I_{D}(\omega) \propto \sum_{\alpha} P_{\alpha}|\vec{\mu}_{\alpha}|^2 \tilde{D}_{\alpha}(\omega),
\end{equation}
where $|\vec{\mu}_{\alpha}|$ is the transition dipole strength of exciton $\alpha$ given by $|\vec{\mu}_{\alpha}|^2 = |\sum_{i}C_{i}^{\alpha} \vec{\mu}_{i}|^2$ and $\vec{\mu}_{i}$ is the transition dipole moment at chromophore site i.

\section*{Results}

\begin{figure}[hbt!]
    \centering
    \includegraphics[width=0.85\linewidth]{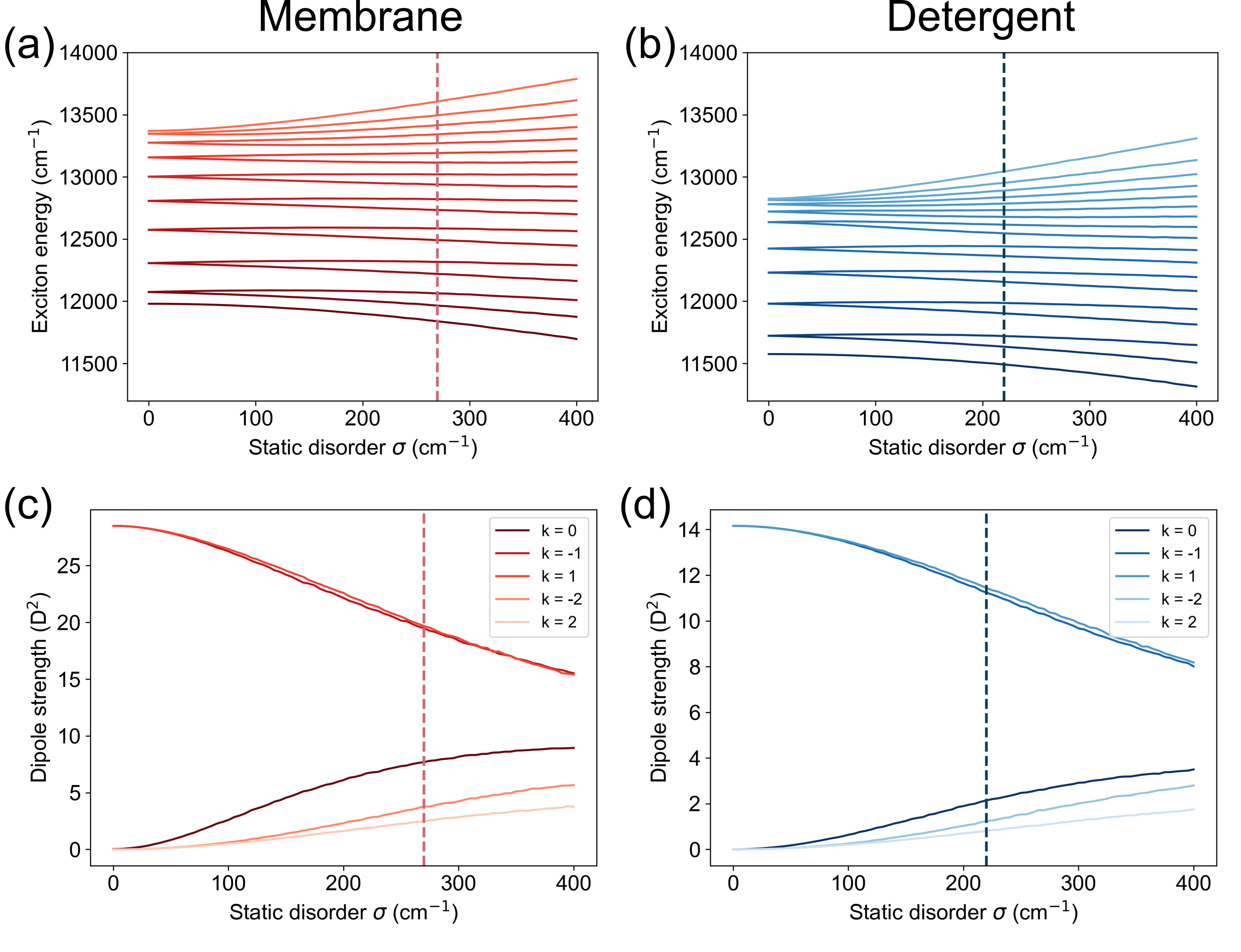}
    \caption{(a-b) Exciton energy levels of the B850 ring as a function of static disorder and (c-d) exciton transition dipole strength for the lowest five levels of the B850 ring averaged over 10,000 realisations for detergent and membrane respectively. The level of static disorder expected in the B850 ring is given by the red dotted line for membrane embedded LH2 and by the blue dotted line for detergent isolated LH2.}
    \label{fig:disorder}
\end{figure}
We begin by examining properties of the excitons that have been well documented by previous theoretical and experimental work on the LH2 and see how they are altered for LH2 embedded in a lipid membrane environment. Motivated by experiment, we focus on comparing POPC membrane LH2 to detergent LH2, but similar conclusions apply to DOPC membrane. We finally compare B800 to B850 transfer rates computed using GFT and HEOM in each environment and determine the main energy transfer pathways that contribute to the transfer rate to see how they change from membrane to detergent.

\subsection*{Exciton energy vs. static disorder}

The B800 and B850 Hamiltonian's were diagonalised to obtain B800 exciton energies and B850 exciton energies respectively. In the absence of static disorder, the B800 exciton manifold consists of one low lying energy level, followed by four pairs of doubly degenerate levels. In the B850 manifold, the lower energy exciton levels have a similar structure consisting of one low energy level followed by four doubly degenerate levels. The higher energy levels, B850* \cite{novoderezhkin2003intra, fidler2013probing}, consist of four doubly degenerate levels followed by a single highest energy level. 

Figure \ref{fig:disorder}(a) and \ref{fig:disorder}(b) gives the exciton energy levels of the B850 ring as a function of static disorder averaged over 10,000 realisations for membrane and detergent LH2 respectively. As static disorder increases the degeneracy of the exciton levels is lifted and the average energy levels begin to diverge. Following the inclusion of static disorder, the k quantum number for each eigenstate is no longer well defined. Here, we use k simply for ease of labelling, with negative k values referring to the lower energy level.

\subsection*{Exciton transition dipole strength vs. static disorder}
Through the interaction of $\vec{\mu}_{\alpha}$ with an electromagnetic field, an optical transition from the ground state to the excited state, or vice versa, is possible. Thus, $|\vec{\mu}_{\alpha}|^2$  can tell us if a transition to a given exciton state is optically allowed, as it defines the strength of the interaction between $\vec{\mu}_{\alpha}$ the electromagnetic field.

Figure \ref{fig:disorder}(c) and \ref{fig:disorder}(d) gives $\vec{\mu}_{\alpha}$ for the five lowest lying levels in the B850 ring for increasing static disorder averaged over 10,000 realisations for membrane and detergent LH2 respectively. Without accounting for static disorder the k$= \pm1$ states of the B850 ring are the only bright states, i.e. almost all of the transition dipole strength in the system is associated with them. As static disorder increases, the transition dipole strength is redistributed to neighbouring exciton states that are close in energy to k$ = \pm1$, namely k$ = 0, \pm2$. The k$ = \pm1$ states still retain a majority of the transition dipole strength when accounting for static disorder, making them most important for energy transfer to the B850 ring via optical transitions.

\begin{figure}[hbt!]
    \centering
    \includegraphics[width=0.95\linewidth]{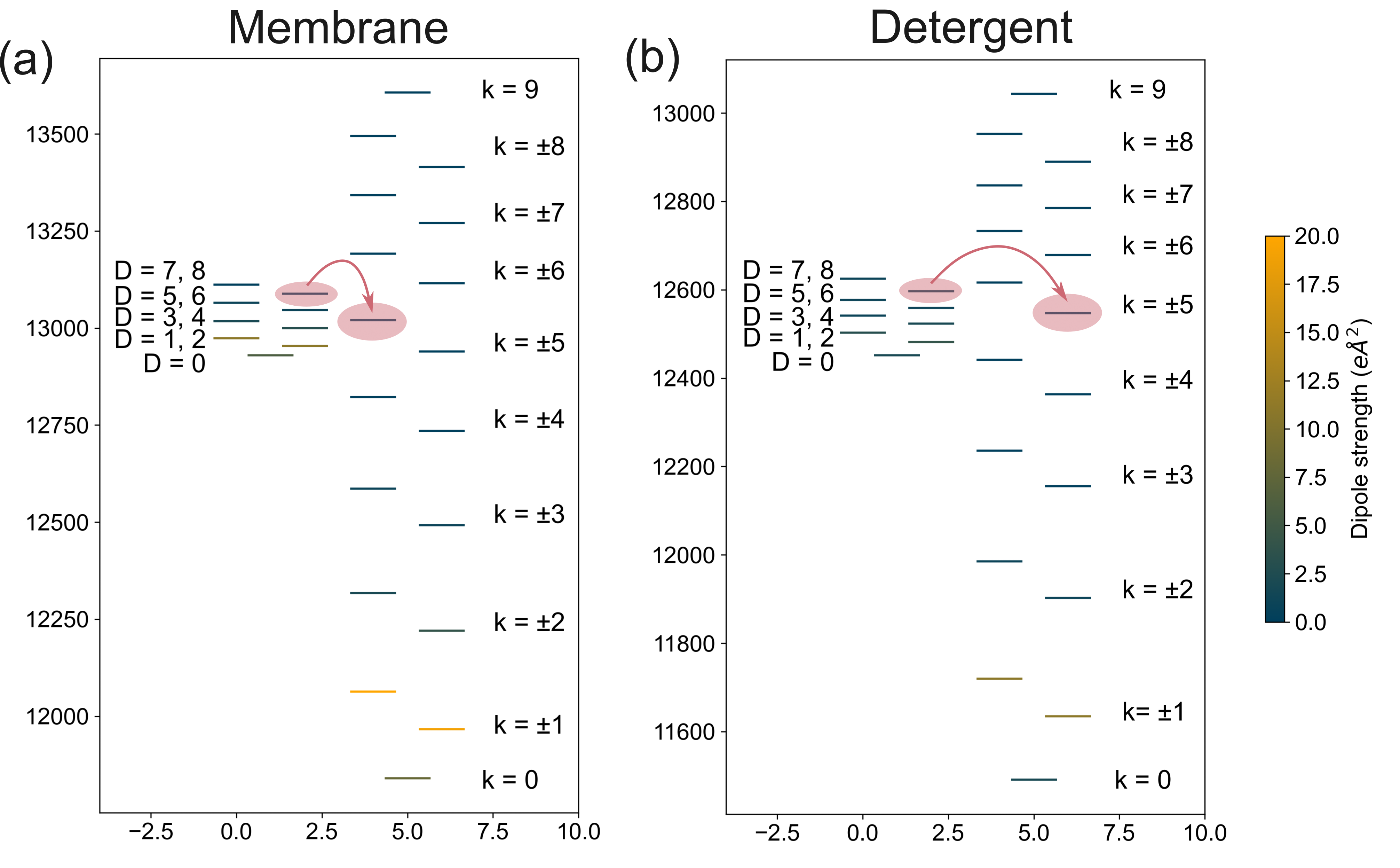}
    \caption{Average positions of the exciton energy levels of the B800 and B850 rings of (a) POPC membrane embedded LH2 and (b) detergent isolated LH2 averaged over 10,000 realisations of static disorder. The disorder expected in each ring for each environment is given in Table \ref{tab:1}. Red arrows indicate the dominating B800 to B850 energy transfer pathways, determined using GFT.}
    \label{fig:energylevels}
\end{figure}

\subsection*{Exciton energy levels and dipole strengths at defined static disorder}

Figure \ref{fig:energylevels} shows the average positions of the B800 and B850 exciton energy levels calculated using 10,000 realisations of static disorder for membrane embedded LH2 and detergent isolated LH2. 

At higher levels of static disorder, energy levels diverge more, therefore we should expect a greater increase in the width of the average exciton manifold where there is high disorder. In the B850 ring static disorder is 50 cm$^{-1}$ higher in membrane so the B850 manifold width increases more in membrane (27\%) than in detergent (24\%). In the B800 ring static disorder is 10 cm$^{-1}$ greater in detergent LH2 and the B800 manifold width increases ~3 times as much in detergent compared to in membrane, suggesting that the B800 ring is more sensitive to static disorder. Weak nearest neighbour electronic couplings in the B800 ring are less than the levels of static disorder in the ring. Changes in the B800 site energies are therefore comparable to the coupling strength between them making the excitonic structure more sensitive to static disorder.

Due to the differences in the average excitonic structure in membrane and detergent LH2, the B800 excitons overlap spectrally with different B850 excitons in each environment, which impacts the key B800 to B850 energy transfer pathways. For the membrane, there is a greater overlap on average between the B800 states and the the dark B850* states, while for detergent the overlap is with lower energy B850 states. The energy transfer pathways that dominate the B800 to B850 transfer in each environment as determined by GFT is shown by the red arrows in Figure \ref{fig:energylevels}. Differences in energy transfer pathways can result in differences in the overall B800 to B850 transfer rate.

\subsection*{Exciton delocalisation}
We can examine differences in the delocalisation of excitons in membrane and detergent LH2 by calculating the $C_{l1}$ for excitons localised on each respective ring, where excitations are understood to be superpositions of excited states localised on single sites. We additionally calculate the IPR of the excitons and compare the two measure of delocalisation.

\begin{table}[hbt!]
    \small
    \centering
    \begin{tabular}{l|ccc|ccc}
    \hline
        \multicolumn{1}{c}{} & \multicolumn{3}{|c|}{$C_{l1}$} & \multicolumn{3}{c}{IPR}\\
    \hline
        \textrm{Exciton} & \textrm{POPC Membrane} & \textrm{DOPC Membrane} & \textrm{Detergent}    & \textrm{POPC Membrane}    & \textrm{DOPC Membrane}& \textrm{Detergent}\\
    \hline
    k = 9   & 4     & 3     & 3     & 3     & 2     & 2     \\
    k = +8  & 5     & 4     & 3     & 3     & 3     & 3     \\  
    k = -8  & 6     & 5     & 4     & 4     & 3     & 3     \\
    k = +7  & 7     & 6     & 5     & 5     & 4     & 3     \\
    k = -7  & 9     & 8     & 6     & 5     & 4     & 4     \\
    k = +6  & 9     & 8     & 7     & 6     & 5     & 4     \\
    k = -6  & 11    & 9     & 8     & 7     & 6     & 5     \\
    k = +5  & 11    & 10    & 9     & 7     & 6     & 5     \\
    k = -5  & 12    & 11    & 10    & 8     & 7     & 6     \\
    average & 8     & 7     & 6     & 5     & 4     & 4     \\
    \hline
    k = +4  & 12    & 11    & 11    & 8     & 7     & 7     \\
    k = -4  & 12    & 11    & 11    & 9     & 8     & 9     \\
    k = +3  & 12    & 12    & 12    & 9     & 8     & 9     \\
    k = -3  & 12    & 11    & 12    & 9     & 8     & 9     \\
    k = +2  & 13    & 12    & 13    & 9     & 8     & 10    \\
    k = -2  & 12    & 11    & 12    & 8     & 8     & 9     \\
    k = +1  & 12    & 12    & 13    & 8     & 8     & 10    \\
    k = -1  & 10    & 10    & 11    & 7     & 6     & 8     \\
    k = 0   & 10    & 9     & 13    & 6     & 5     & 9     \\
    average & 12    & 11    & 12    & 8     & 7     & 9     \\
    \hline
    average & 10    & 9     & 9     & 7     & 6     & 6     \\
    
    \hline
    \end{tabular}
\caption{The $l1$ norm of coherence and inverse participation ratio of the B850 excitons in POPC membrane embedded LH2, DOPC membrane embedded LH2 and detergent isolated LH2 averaged over 10,000 realisations of disorder listed from highest to lowest energy. A line dividing the states in half separates what we describe as the high energy manifold from the low energy manifold.}
\label{tab:2a}
\end{table}

\begin{table}[hbt!]
    \small
    \centering
    \begin{tabular}{l|ccc|ccc}
    \hline
        \multicolumn{1}{c}{} & \multicolumn{3}{|c|}{$C_{l1}$} & \multicolumn{3}{c}{IPR}\\
    \hline
        \textrm{Exciton} & \textrm{POPC Membrane} & \textrm{DOPC Membrane} & \textrm{Detergent}    & \textrm{POPC Membrane}    & \textrm{DOPC Membrane}& \textrm{Detergent}\\
    \hline
    D = 8   & 3     & 4     & 2     & 3     & 3     & 2     \\
    D = 7   & 5     & 5     & 2     & 4     & 4     & 2     \\  
    D = 6   & 5     & 5     & 3     & 4     & 4     & 2     \\
    D = 5   & 6     & 6     & 3     & 5     & 5     & 3     \\
    D = 4   & 6     & 6     & 4     & 5     & 5     & 3     \\
    D = 3   & 5     & 5     & 4     & 4     & 5     & 3     \\
    D = 2   & 6     & 6     & 4     & 4     & 4     & 3     \\
    D = 1   & 4     & 4     & 3     & 3     & 3     & 2    \\
    D = 0   & 4     & 4     & 3     & 3     & 3     & 2     \\
    \hline
    average & 5     & 5     & 3     & 4     & 4     & 2     \\
    \hline
    \end{tabular}
\caption{The $l1$ norm of coherence and inverse participation ratio of the B800 excitons in POPC membrane embedded LH2, DOPC membrane embedded LH2 and detergent isolated LH2 averaged over 10,000 realisations of disorder listed from highest to lowest energy.}
\label{tab:2b}
\end{table}

\begin{figure}[hbt!]
    \centering
    \includegraphics[width=0.9\linewidth]{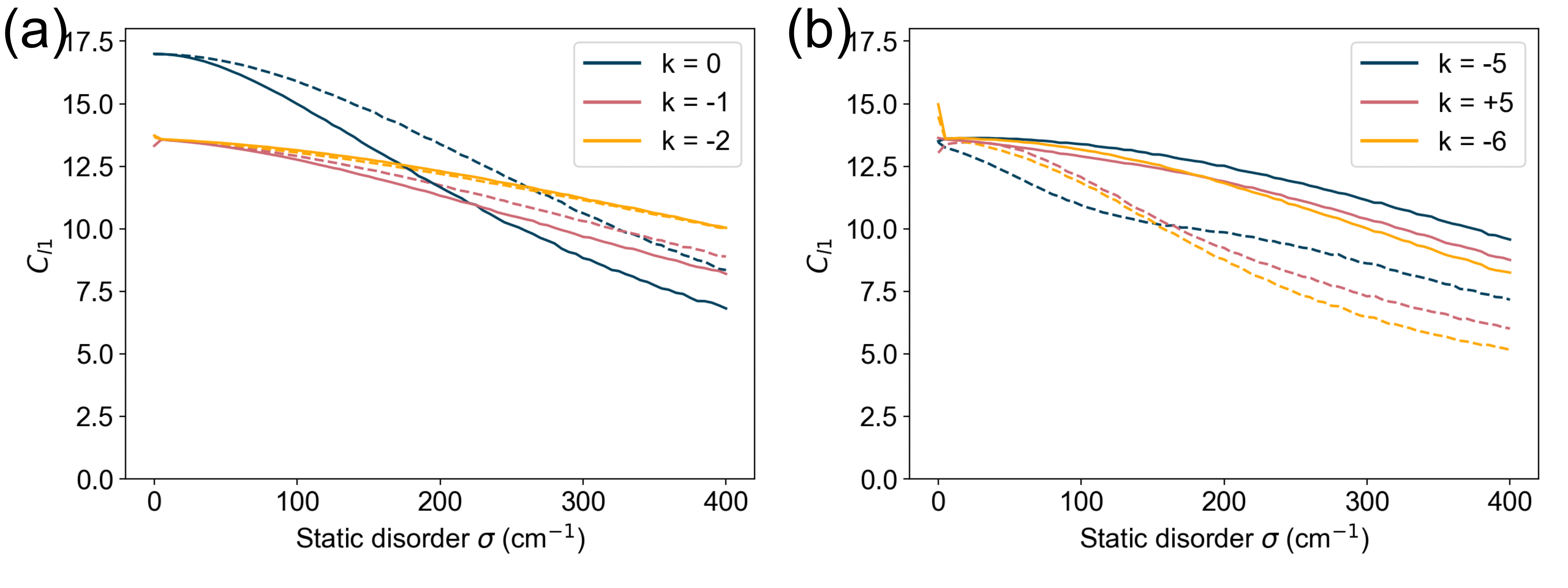}
    \caption{The average $l1$ norm of coherence of three levels selected from the (A) lower energy exciton manifold and (B) high energy exciton manifold of the B850 ring as a function of static disorder. Solid lines are for membrane-embedded LH2 while dotted lines are for the detergent-isolated LH2. The low energy excitons show a reduced delocalisation in the membrane, while the high energy excitons have increased delocalised in the membrane. An average over 10,000 realisations of static disorder was used to compute $C_{l1}$.}
    \label{fig:deloc}
\end{figure}

At zero static disorder, excitons have the same IPR in all environments, apart from a small 6\% increase in POPC membrane and DOPC membrane for four B850 levels, k$=\pm4$ and k=$\pm5$ relative to the same excitons in detergent.
Noticeable differences start to emerge when the IPR is calculated at the level of static disorder expected in each ring. Table \ref{tab:2a} and \ref{tab:2b} gives $C_{l1}$ and the IPR averaged over 10,000 realisations of static disorder for B850 and B800 excitons respectively. An overall decrease in the delocalisation is seen across all excitons. This is because static disorder creates random shifts in the electronic parameters of the chromophores such that their site energies are no longer identical, reducing the symmetry of the system which tends to localise the excitons. However in each environment, the localising effect of static disorder perturbs each exciton differently.

$C_{l1}$ is a proper measure of coherence based on distance measures, hence provides a more reliable value to compare the delocalisation of different states. For example, take the B850 states k = -3 and k = +2, the IPR is equal for these states, yet $C_{l1}$ reveals that they have different delocalisation. For other states the IPR predicts different delocalisation when $C_{l1}$ shows that those states have identical delocalisation. Thus the IPR can be misleading when a comparison of state delocalisation is desired. 

Comparing the average $C_{l1}$ of the B800 excitons, ($\overline{C_{l1}(\rho^{\alpha})} = \frac{1}{N}\sum_{\alpha\in B850}^{N} C_{l1}(\rho^{\alpha})$), excitons are the least delocalised in detergent compared to the membrane environments, as expected due to the higher level of static disorder and weaker electronic couplings in the B800 ring in detergent. Comparing the average $C_{l1}$ for all B850 states in each environment we find a larger average delocalisation of states in POPC membrane than in detergent and DOPC membrane. This seems to arise primarily from the high energy dark states of the B850 ring having increased delocalisation as the lower energy bright states tend to have reduced delocalisation compared to detergent LH2. 

Figure \ref{fig:deloc} compares $C_{l1}$ with increasing static disorder for three low energy and three high energy exciton states of the B850 ring in POPC membrane and detergent. We find that the order of the exciton delocalisation changes depending on the level of static disorder. Amongst the high energy levels in POPC membrane, the k = -6 level is more delocalised than the k = +5 level at static disorder below 200 cm$^{-1}$ (Figure \ref{fig:deloc}b). Above 200 cm$^{-1}$ this is reversed and the k=+5 level is more delocalised. In POPC membrane, there is a reduction in delocalisation of some states in the low energy manifold (k = 0 to k = +4) relative to the detergent states as expected due to static disorder being greater in the B850 ring of the membrane. Some states in the high energy manifold (k = -5 to k = 9) display increased delocalisation in POPC membrane, a result of stronger electronic couplings in the B850 ring, which results in the high and low energy B850 excitons having a more comparable delocalisation than in detergent. To quantify this, The difference between the average $C_{l1}$ of the high energy manifold and low energy manifold is 4 for both membrane models and 6 for the detergent model. Thus in detergent LH2, there is a clear distinction in the delocalisation between the lower energy exciton manifold and the high energy manifold which is less pronounced in membrane environments.

These calculations suggest that the membrane tends to preserve the symmetry of the excitonic structure of the B850 ring by tuning the delocalisation of the high and low energy exciton manifolds thereby enhancing a quantum feature of the system. Since an excitonic description is required to accurately predict energy transfer rates in LH2, changes in exciton delocalisation could manifest as changes in the energy transfer pathways of an excitation \cite{chachisvilis1997excitons}, altering coherent dynamics in LH2 when embedded in the membrane. As the system evolves in time, exciton delocalisation can change due to environmental interactions \cite{fassioli2014photosynthetic}. More sophisticated measures can help verify if these differences in delocalisation from detergent to membrane persist over the interring energy transfer timescales.

\subsection*{Theoretical linear spectra}
\begin{figure}[hbt!]
    \centering
    \includegraphics[width=1.0\linewidth]{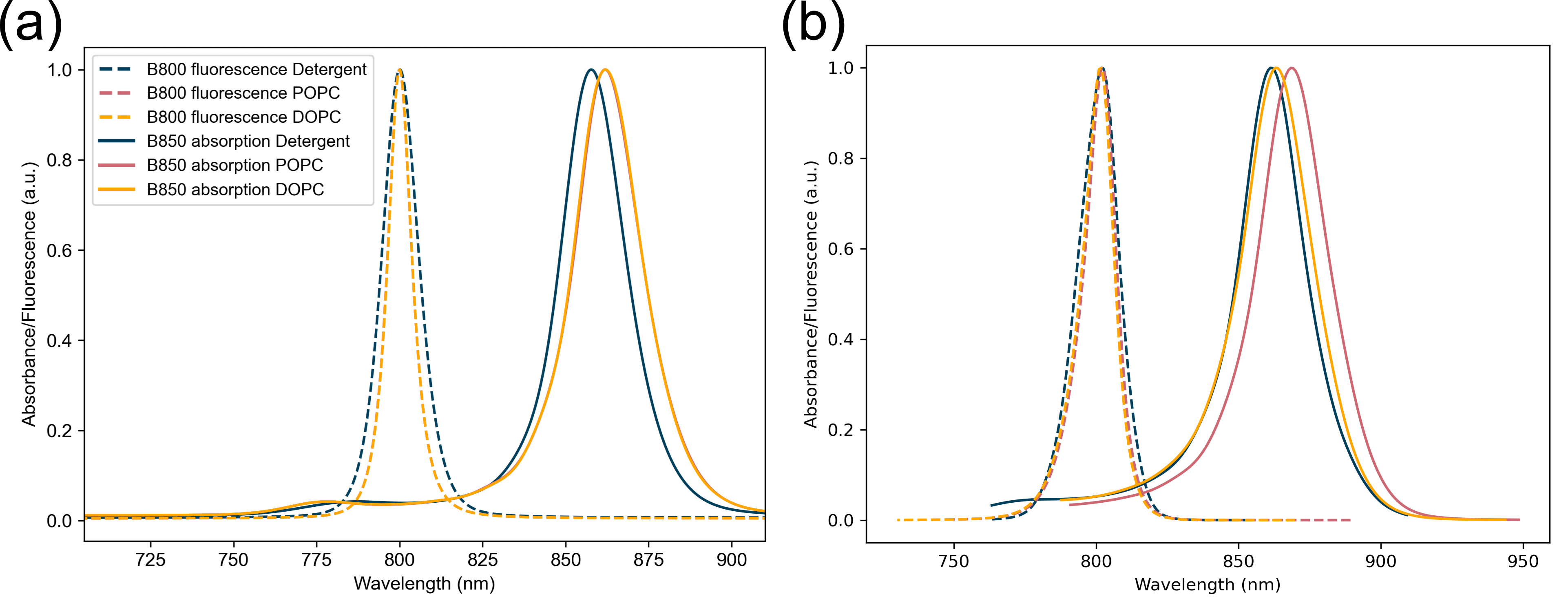}
    \caption{Theoretical linear spectra of the individual B800 and B850 rings computed using (a) lineshape theory as outlined in the \nameref{Sec:GFTTheory} section, averaged over 2000 realisations of disorder, and (b) using HEOM as outlined in the \nameref{Sec:HEOMTheory} section, averaged over 500 realisations of disorder. Convergence of the HEOM spectra is obtained when the hierarchy is truncated at tier 3 with the exception of the B800 fluorescence which is truncated at tier 4.}
    \label{fig:spectra}
\end{figure}

One of the key differences seen in experiments comparing detergent isolated and membrane embedded LH2 is the redshift of the B850 band in the linear absorption spectra of the LH2 \cite{freiberg2012comparative, agarwal2002nature, ogren2018impact}. In Figure \ref{fig:spectra}(a) we have computed the B850 absorption and B800 fluorescence for membrane and detergent LH2 using the same lineshape theory that is used to compute energy transfer rates in GFT and HEOM.

The redshift of the B850 absorption peak is obtained at both levels of theory, although it is exaggerated in the lineshape theory spectra. While the lineshape spectra predicts the same redshift for both POPC and DOPC membrane, HEOM is able to resolve differences between the two lipid environments. Isolated BChl a's absorb at 800 nm, but when they come together to form the B850 ring, interchromophore electronic interactions shift the 800 nm absorption peak to 850 nm \cite{sauer1996structure}. Therefore, the observed shift in the B850 absorption spectrum between POPC membrane and detergent likely arises due to stronger interchromophore couplings and consequently increased delocalisation of excitons in the membrane. In DOPC the reduced redshift compared to POPC is possibly related to the intradimer B850 couplings being weaker that in detergent while interdimer couplings are stronger. 

The redshift of the B850 band reduces the spectral overlap of the B800 fluorescence and B850 absorption bands which would imply slower B800 to B850 energy transfer times in the membrane. Since measured energy transfer rates are larger in the membrane, this signals that the increased delocalisation of the B850 excitons compensates for the slightly reduced overlap. Of the B850 excitons, the dark states show the greatest increase in delocalisation in the membrane, hence playing an important role in the energy transfer. 

\subsection*{B800 to B850 transfer rate distribution}

\begin{figure}[hbt!]
    \centering
    \includegraphics[width=0.9\linewidth]{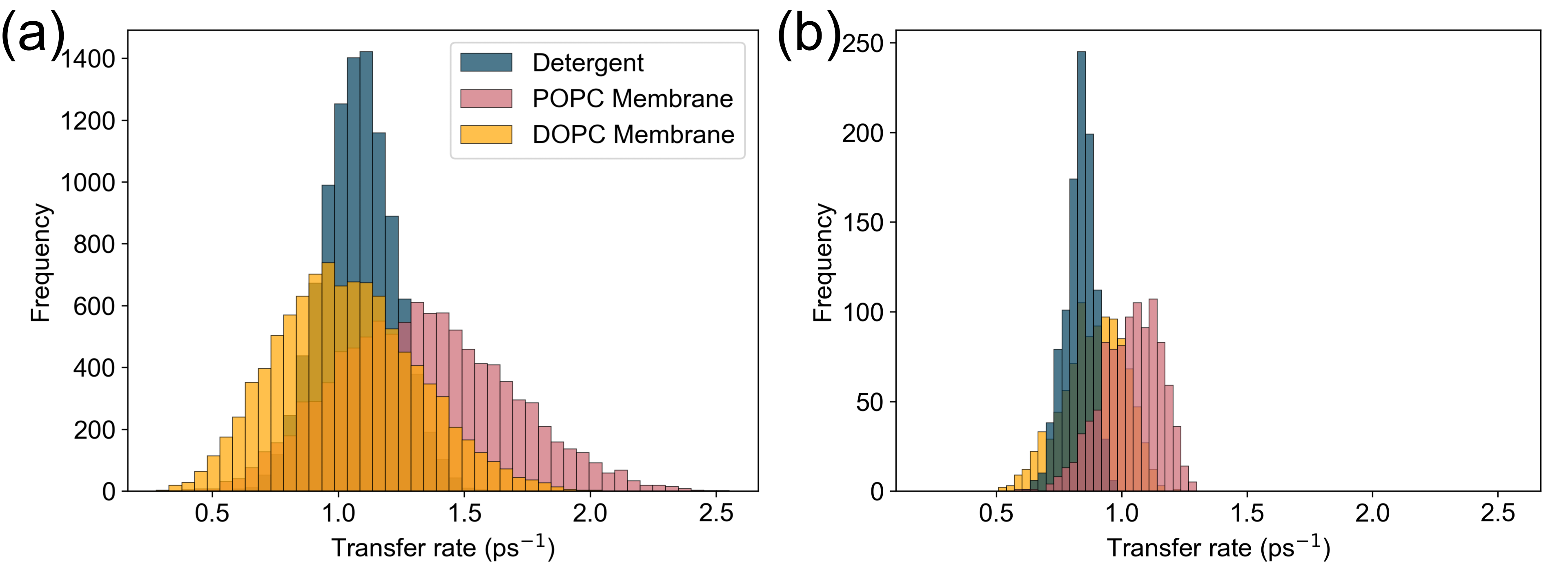}
    \caption{(a) Distribution of 10,000 realisations of the B800 to B850 energy transfer rate calculated using GFT for detergent isolated LH2, and two different lipid compositions of membrane embedded LH2, DOPC and POPC. Average transfer rates are 1.08 ps$^{-1}$, 1.04 ps$^{-1}$ and 1.34 ps$^{-1}$ respectively. Standard deviations are 0.14 ps$^{-1}$, 0.28 ps$^{-1}$ and 0.34 ps$^{-1}$ respectively.(b)  Distribution of 2000 realisations of the B800 to B850 energy transfer rate calculated using the fitting population equation Eq. (\ref{eq:pauli}) with rates obtained through HEOM. Average transfer rates are 0.83 ps$^{-1}$, 0.89 ps$^{-1}$ and 1.04 ps$^{-1}$ respectively. Standard deviations are 0.06 ps$^{-1}$, 0.12 ps$^{-1}$ and 0.12 ps$^{-1}$ respectively.}
    \label{fig:transferrate}
\end{figure}
The B800 to B850 energy transfer rate was calculated for membrane embedded LH2 and detergent isolated LH2 using GFT (Eq. (\ref{eq:gft})). 10,000 realisations of static disorder were used for each environment and the distribution of the transfer rate over these realisations are shown in Figure \ref{fig:transferrate}(a). In qualitative agreement with experimental work, the B800 to B850 transfer rate in POPC membrane LH2 has a faster average of 1.34 ps$^{-1}$ (746 fs) compared to the rate in detergent where the average is 1.08 ps$^{-1}$ (925 fs). To corroborate rates obtained using GFT, estimates of the B800 to B850 energy transfer rates for 2000 realisations of disorder have been computed using HEOM per the procedure outlined in the \nameref{Sec:HEOMTheory} section and are shown as a histogram in figure \ref{fig:transferrate}(b). In agreement with GFT rates, average energy transfer is found to be faster in POPC membrane at 1.04 ps$^{-1}$ (962 fs) compared to detergent at 0.83 ps$^{-1}$ (1.2 ps). Average rates obtained using HEOM are faster than those determined by GFT, which is likely a result of mapping the B800 to B850 transfer process to a one-step process while GFT considers multiple simultaneous B800 to B850 exciton transfer processes. Despite this discrepancy, there is still qualitative agreement between both the exact and perturbative rates indicating that GFT is able to capture some differences between detergent and membrane LH2.

To understand the microscopic origin of the increased energy transfer rate in membrane LH2, we have computed the B800 to B850 energy transfer rate for LH2 embedded in a membrane composed of a different lipid species, 1,2-dioleoyl-\textit{sn}-glycero-3-phosphocholine (DOPC) \cite{ramos2019molecular}. In a DOPC membrane, the average B800 to B850 transfer rate is slower than in POPC membrane. This suggests that lipid species can influence energy transfer rates within the LH2. There is evidence that changes in the lipid composition of the bilayer impacts lateral pressure and electric field profiles, which leads to changes in the conformational equilibrium of membrane proteins \cite{cantor1997lateral, ding2015effects}. Phospholipids in the bacterial membrane that hosts the LH2 could present lateral pressure and electric field profiles that differ from the detergent environment that may alter the electronic structure, and therefore function, of the LH2 \cite{ogren2018impact}.

Despite being a perturbative theory GFT can still distinguish differences in B800 to B850 energy transfer within the LH2 in different lipid compositions. However, GFT rates become unreliable when a drastic change in environment from lipid to detergent is made. GFT predicts faster transfer in detergent than DOPC membrane, contradicting experimental results that find faster transfer in a membrane environment. Meanwhile HEOM derived rates predict slower transfer in detergent as expected from experiment. The discrepancy between the GFT and HEOM rates is likely linked to two approximations made in GFT that result in contributions from non-equilibrium and coherent effects being neglected. Firstly, GFT assumes the initial state is limited to being localised on the B800 ring and that the excitation then “hops” to a state localised on the B850 ring. HEOM instead allows for the initial state to evolve coherently from the B800 ring to the B850 ring, thus allowing for a state to be delocalised over both rings during transfer. Secondly, GFT assumes the initial B800 state begins and remains in thermal equilibrium. The HEOM derived rate accounts for non-equilibrium effects during transfer as the interaction with the environment results in the initial state evolving and shifting out of equilibrium. Thus the non-equilibrium and coherent contributions to the B800 to B850 energy transfer rate are key to predict differences in the LH2’s function in detergent and membrane.

The broader distribution of B800 to B850 energy transfer rates for membrane LH2 shows that the transfer rate varies by more across each complex in an ensemble in membrane than in detergent. The larger standard deviation of the energy transfer rate in membrane implies that the energy transfer process is less precise. The negative relationship between the energy transfer rate and the standard deviation of the energy transfer rate for LH2 in different environments suggests the possibility of a speed-accuracy trade-off within the LH2 \cite{garland2014trade}. Trade-offs exist on a molecular level, with processes like protein synthesis prioritising speed over fidelity \cite{johansson2012genetic}. Such a trade-off is the result of the biological system possessing a trait that cannot increase without the decrease of another trait. For energy transfer in the LH2, the traits involved may be related to lipid properties determined by the lipid composition of the bacterial membrane. However, while a negative relationship is a prerequisite for a trade-off, it is not sufficient and laboratory evolution experiments are required to identify whether a trade-off exists.

\subsubsection*{Dominating exciton transfer pathways}
\begin{figure}[hbt!]
    \centering
    \includegraphics[width=0.99\linewidth]{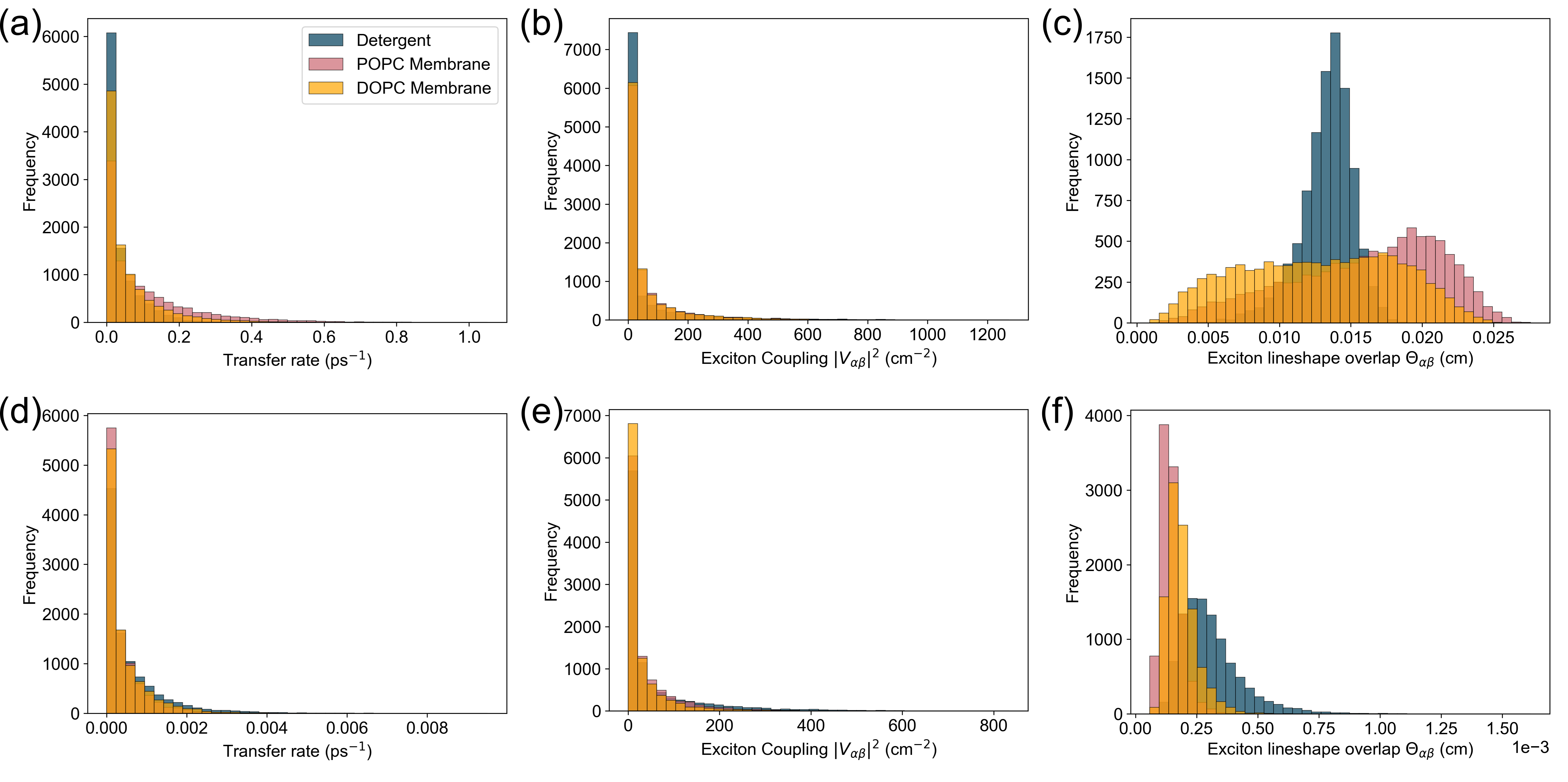}
    \caption{Exciton properties for (a-c) exciton pairs that dominate the B800 to B850 energy transfer and (d-e) exciton pairs that have a small contribution to the B800 to B850 energy transfer rate (non-dominating). Distribution of 10,000 realisations of the exciton energy transfer rate, the exciton coupling and the exciton spectral overlap between a B800 exciton to a B850 exciton Transfer rates are calculated using Eq. (\ref{eq:forstertheory}) and weighted by the thermal occupation of the donor B800 state. The dominating excitons are D = 7 to k = +5 for POPC and DOPC membrane LH2 and D = 7 to k= -5 for detergent LH2. The non-dominating excitons are D = 1 to k = -2.}
    \label{fig:dominating}
\end{figure}

To understand the microscopic differences underlying the change in the B800 to B850 energy transfer rate from detergent to membrane, the dominating exciton energy transfer pathways were determined in each environment. We identify the important transfer pathways as being between excitons with the fastest average exciton transfer rate, as they have the greatest influence on the average B800 to B850 transfer rate. The exciton transfer rate (Eq. \ref{eq:forstertheory}) was determined between all combinations of donor B800 and acceptor B850 excitons and averaged over 10,000 realisations of disorder. For each realisation, the exciton transfer rate is weighted by the thermal occupation of the donor state, in order to correctly weight its contribution to the average B800 to B850 transfer rate (Eq. (\ref{eq:gft})). 

The important exciton energy transfer pathways for B800 to B850 transfer in each environment are given in Table \ref{tab:3}. In all environments, the dominating energy transfer pathway is via dark B850 states, although the specific excitons are slightly different. This is due to differences in the spectral overlap of the excitons in each environment which can be seen by the different relative positions of the B800 and B850 average energy levels in Figure \ref{fig:energylevels}. In the POPC membrane, the B800 levels overlap with higher energy B850 states than in detergent. As a result, the important energy transfer pathways for B800 to B850 transfer are altered compared to detergent LH2. Despite the B800 to B850 transfer rate being slowest in DOPC membrane, the dominating pathway in DOPC membrane is faster than the pathway in detergent. We find that in detergent multiple pathways with moderate transfer rates (~0.03 ps$^{-1}$) exist from low energy B800 levels, while in DOPC membrane transfer from those levels is much slower (>0.01 ps$^{-1}$). Thus GFT predicts the overall B800 to B850 rate to be faster in detergent than DOPC due to the increased number of available pathways for an excitation to take. 

\begin{table}[hbt!]
    \small
    \centering
    \begin{tabular}{lcccccccc}
    \hline
        \textrm{LH2 environment} & \textrm{B800 state}    & \textrm{B850 state}    & \textrm{$P_{\alpha}k_{\alpha\beta}$ (ps$^{-1}$)}  & \textrm{$k_{\alpha\beta}$ (ps$^{-1}$)} & \textrm{$|V_{\alpha\beta}|^2$ (cm$^{-1}$)} & \textrm{$\Theta_{\alpha\beta}$} & \textrm{$C_{l1}^{\mathrm{B}850}$} & \textrm{$C_{l1}^{\mathrm{B}800}$}\\
    \hline
    Membrane POPC   & D = 7     & k = +5    & 0.11     & 1.39   & 66    & 0.016     & 11     & 5     \\
    \hline
    Membrane DOPC   & D = 7     & k = +5    & 0.06     & 0.77   & 63    & 0.013     & 10     & 5     \\
    \hline
    Detergent       & D = 7     & k = -5    & 0.04     & 0.43   & 56    & 0.013     & 10     & 2     \\
    \hline
    \end{tabular}
\caption{The B800-B850 exciton pair that provides the fastest energy transfer pathway from B800 to B850 in each environment, their exciton transfer rate, exciton coupling, spectral overlap and the $l1$ norm of coherence of the donor and acceptor states averaged over 10,000 realisations of disorder.}
\label{tab:3}
\end{table}

\begin{table}[hbt!]
    \small
    \centering
    \begin{tabular}{lcccccccc}
    \hline
        \textrm{LH2 environment} & \textrm{B800 state}    & \textrm{B850 state}    & \textrm{$P_{\alpha}k_{\alpha\beta}$ (ps$^{-1}$)}   & \textrm{$k_{\alpha\beta}$ (ps$^{-1}$)} & \textrm{$|V_{\alpha\beta}|^2$ (cm$^{-1}$)} & \textrm{$\Theta_{\alpha\beta}$} &  \textrm{$C_{l1}^{\mathrm{B}850}$} & \textrm{$C_{l1}^{\mathrm{B}800}$}\\
    \hline
    Membrane POPC   & D = 1     & k = -2    & 0.00035     & 0.0024   & 38    & 1.4   & 12     & 4   \\
    \hline
    Membrane DOPC   & D = 1     & k = -2    & 0.00043     & 0.0031   & 28    & 1.9   & 11     & 4   \\
    \hline
    Detergent       & D = 1     & k = -2    & 0.0006     & 0.0043   & 57    & 3.1   & 12     & 3   \\
    \hline
    \end{tabular}
\caption{A B800-B850 exciton pair that has slow exciton energy transfer from B800 to B850, their exciton transfer rate, exciton coupling, spectral overlap and the $l1$ norm of coherence of the donor and acceptor states averaged over 10,000 realisations of disorder.}
\label{tab:4}
\end{table}

Previously, we showed that on average the delocalisation of B850 excitons is greater in membrane LH2. The delocalisation of the B850 excitons that are key energy acceptors in B800 to B850 energy transfer is of greater importance, as it allows us to assess if the change in delocalisation is relevant to the change in energy transfer that we see in membrane LH2. We find that in the membrane, the important B850 excitons are on average more delocalised ($C_{l1}$ = 12) than the equivalent in detergent ($C_{l1}$ = 11). This suggests that coherent dynamics in LH2 may be altered when embedded in the membrane. Further investigation would require the use of HEOM as GFT does not provide information on coherent dynamics.

The exciton transfer rate entering GFT depends on the exciton coupling strength squared and the spectral overlap between the exciton line shapes. Looking at how these properties change from detergent to membrane can help identify which specific differences in membrane contribute to an increased average transfer rate and broader distribution. The distribution of 10,000 realisations of the exciton transfer rate, the exciton coupling and the exciton spectral overlap was determined for the dominating pathway in each environment and is given in Figure \ref{fig:dominating}(a-c) with average values listed in Table \ref{tab:3}. For comparison, the same exciton properties are given in \ref{fig:dominating}(d-f) for states D = 1 to k = -2, a pair of states that have a slow exciton transfer rate and are therefore considered non-dominating. Average values for the non-dominating excitons are given in Table \ref{tab:4}.

For the dominating excitons, while the average spectral overlap is comparable in all three environments, the average exciton coupling is strongest in POPC membrane. The exciton transfer rate scales with the exciton coupling such that the fastest transfer rate is between the POPC membrane donor and acceptor pair, indicating that the dominant energy transfer pathway is mostly dependent on the exciton coupling strength. The exciton coupling strength scales with the electronic coupling between nearest neighbour B800 and B850 chromophore sites (Table \ref{tab:1}) suggesting that stronger interchromophore electronic coupling is the main factor contributing to the faster energy transfer rate in membrane. Stronger electronic coupling between B800 and B850 chromophore transition dipole moments could arise from a change in the direction of the dipoles or a reduced distance between them. By comparing the position coordinates of the B850 chromophores in each model, neighbouring B850 chromophores are slightly closer in the membrane LH2 model, such that stronger interchromophore electronic couplings would be expected. 

The exciton coupling is additionally dependent on the exciton delocalisation scaled by the electronic coupling (Eq. \ref{eq:excitoncoupling}). Increased delocalisation of the excitons can contribute to the stronger interaction between excitons by spreading an excitation over a greater number of electronically interacting sites, however, the strength of the interaction between those sites is also important. Although the acceptor B850 exciton is delocalised similarly in DOPC membrane and in detergent, the donor B800 exciton is more delocalised and the electronic coupling between sites is stronger in DOPC membrane, resulting in a stronger exciton coupling and faster exciton transfer rate. This suggests that while the B850 ring is more delocalised than the B800 ring, the exciton delocalisation in B800 is still important for energy transfer.

\begin{figure}[hbt!]
    \centering
    \includegraphics[width=0.6\linewidth]{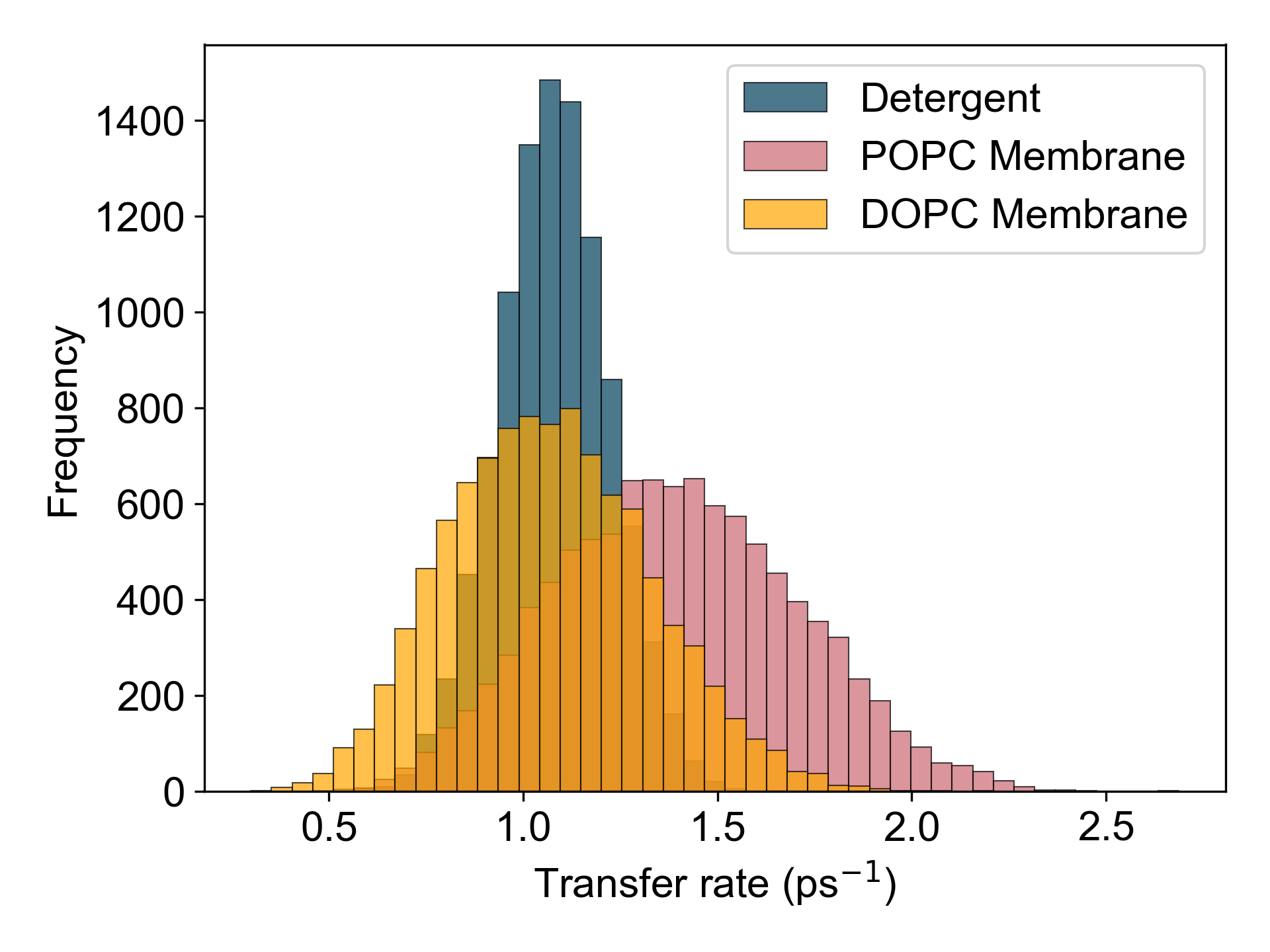}
    \caption{ Distribution of 10,000 realisations of the B800 to B850 energy transfer rate calculated using GFT for detergent isolated LH2, and two different lipid compositions of membrane embedded LH2, DOPC and POPC. The same static disorder parameters were used to calculate rates in all three environments. Average transfer rates are 1.08 ps$^{-1}$, 1.07 ps$^{-1}$ and 1.40 ps$^{-1}$ respectively. }
    \label{fig:samedisorder}
\end{figure}

\begin{figure}[hbt!]
    \centering
    \includegraphics[width=0.6\linewidth]{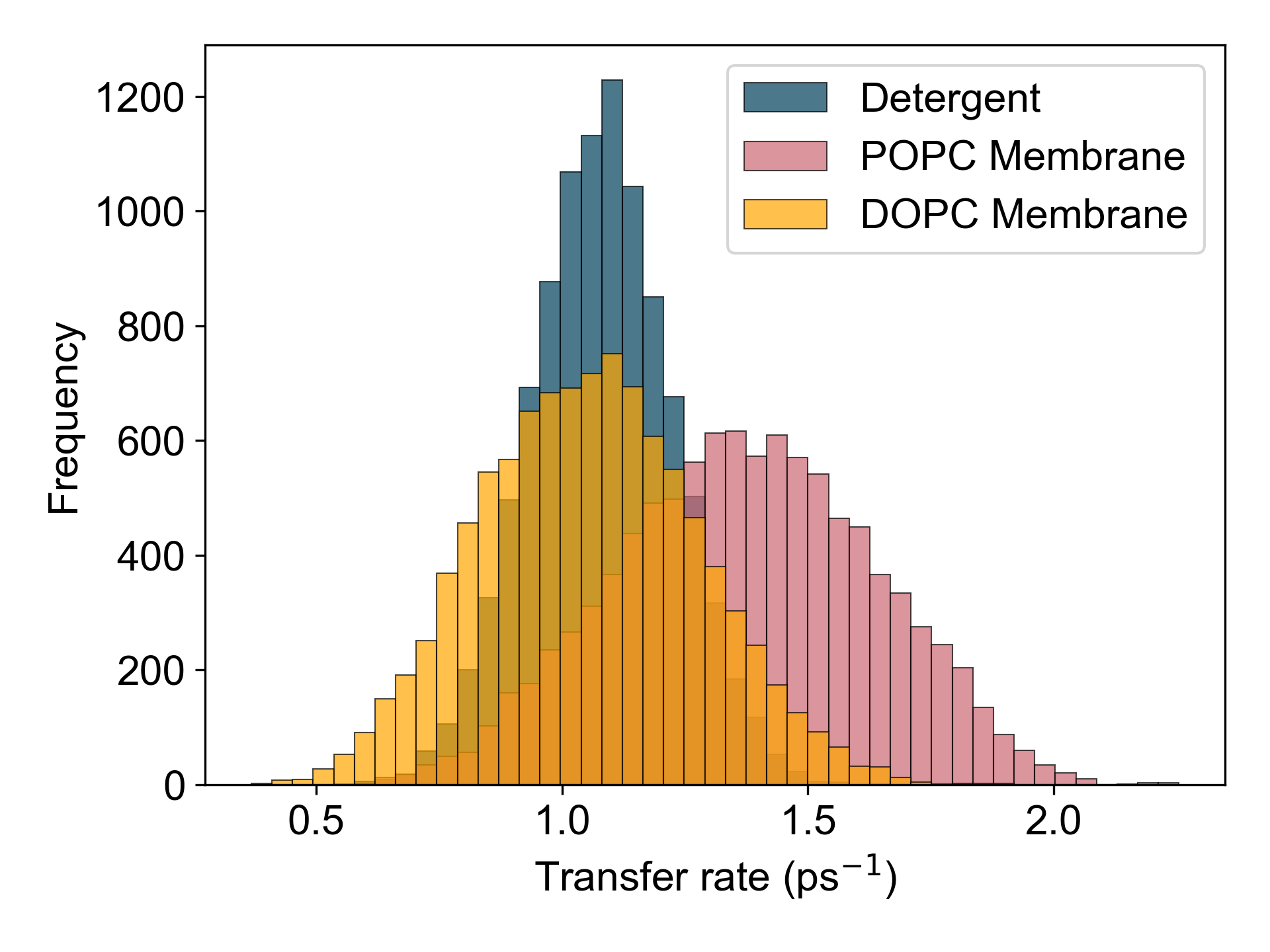}
    \caption{(a) Distribution of 10,000 realisations of the B800 to B850 energy transfer rate calculated using GFT for detergent isolated LH2, and two different lipid compositions of membrane embedded LH2, DOPC and POPC. The same spectral density was used for all three environments. Average transfer rates are 1.08 ps$^{-1}$, 1.05 ps$^{-1}$ and 1.37 ps$^{-1}$ respectively.}
    \label{fig:samesspectraldensity}
\end{figure}

\begin{figure}[hbt!]
    \centering
    \includegraphics[width=0.6\linewidth]{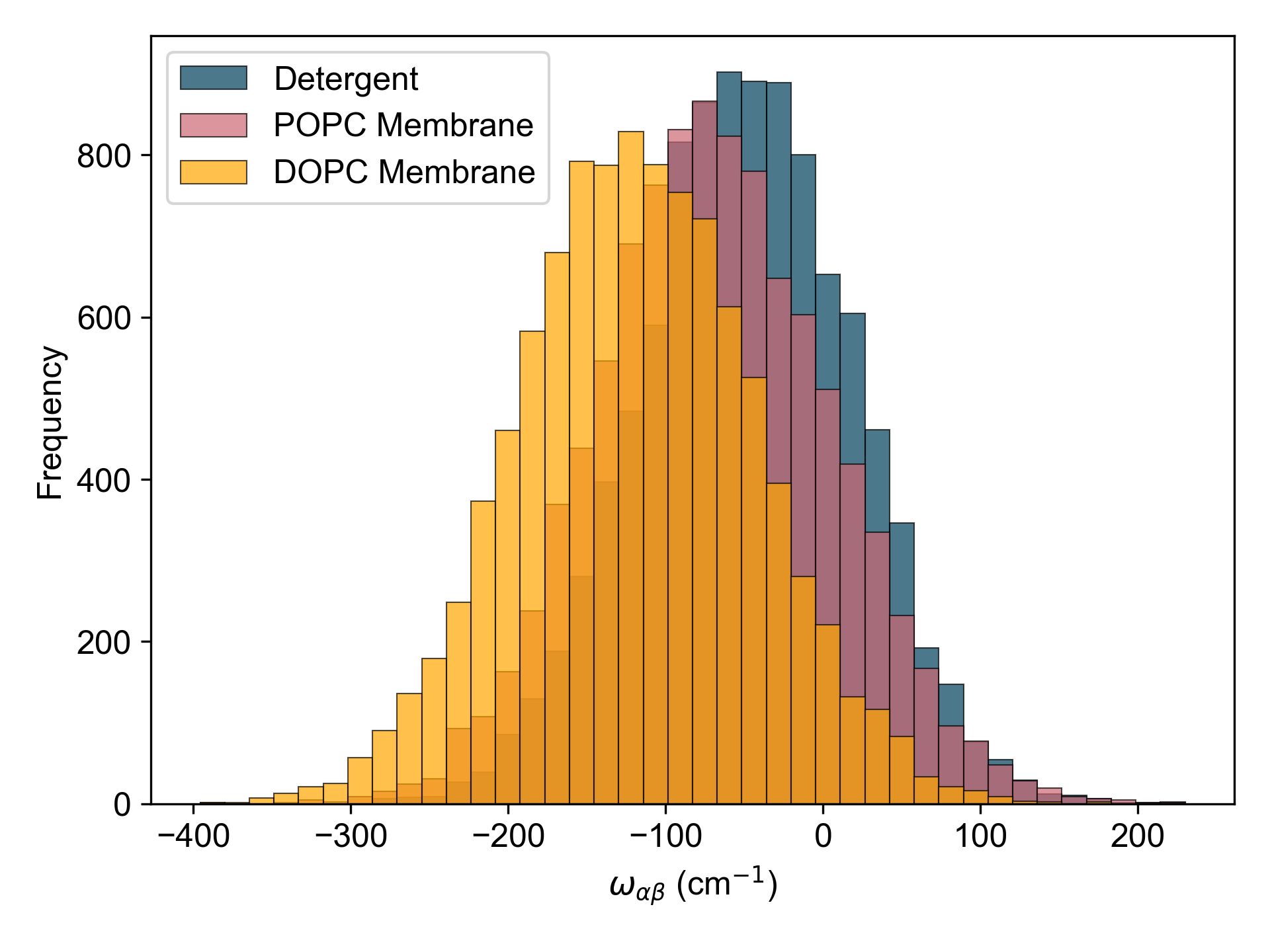}
    \caption{Distribution of 10,000 realisations of the energy gap between the B800 and B850 exciton forming the dominant exciton energy transfer pathway in detergent isolated LH2, LH2 embedded in a DOPC membrane and in a POPC membrane. Despite different levels of static disorder in membrane and detergent environments, the distribution of the energy gap of the dominant pathway in B800 to B850 transfer remains similar in each environment.}
    \label{fig:energygap}
\end{figure}

While the form of the distribution of the exciton coupling is similar in each environment, the distribution of the spectral overlap (Figure \ref{fig:dominating}(c)) is where differences emerge. The spectral overlap distribution peaks sharply for the detergent but is broad and flat for POPC membrane and DOPC membrane which contributes to a greater variation in the exciton transfer rate in membrane. At first, it may seem that this is a consequence of higher static disorder in the membrane which produces random shifts in the relative positions of the donor and acceptor energy levels hence resulting in a greater variation of the energy gap between donor and acceptor. However the distribution of the donor-acceptor energy gap $\omega_{\alpha\beta}$ is similar in all three environments (Figure \ref{fig:energygap}). Additionally, the B800 to B850 transfer rate computed using the same level of static disorder in each ring for detergent and membrane environments still produces a broader distribution and faster average rate for the POPC membrane, suggesting that higher levels of static disorder in the membrane are unlikely to be the cause of the broadened distribution (Figure \ref{fig:samedisorder}). The exciton spectral overlap also depends on the exciton environmental parameters $\lambda_{\alpha\alpha\alpha\alpha}$ and $g_{\alpha\alpha\alpha\alpha}(t)$ which are determined by exciton delocalisation scaled by the site environmental parameters $\lambda_{i}$ and $g_{i}$ via $\lambda_{\alpha\alpha\alpha\alpha} \propto \lambda_{i}/\mathrm{IPR}$. We have computed the B800 to B850 energy transfer rate using the same environmental parameters for both membrane and detergent models and found that a broader distribution and faster rate in membrane LH2 still holds (Figure \ref{fig:samesspectraldensity}). This points to the change in electronic properties being of high importance to the changes in energy transfer in membrane LH2.

For the dominant exciton transfer pathways, we see that the excitonic coupling is the key factor determining the exciton energy transfer rate. The excitonic coupling is dependent on both the electronic coupling between sites in the B800 ring to sites in the B850 ring and the delocalisation of the B800 and B850 exciton in question. A figure of merit that captures both these electronic properties is given by the inter-ring electronic coupling scaled by the geometric average delocalisation of the important B800 and B850 exciton pair, $\sqrt{\overline{C_{l1}^{\mathrm{B}800}}\ \overline{C_{l1}^{\mathrm{B}850}}}V_{B800, \alpha_{2}}$. While $V_{B800, \alpha_{2}}$ provides information on the interactions between the rings, $\sqrt{\overline{C_{l1}^{\mathrm{B}800}}\ \overline{C_{l1}^{\mathrm{B}850}}}$ is a result of the site energies and electronic couplings within each ring. Table \ref{tab:5} lists this figure of merit for the dominating pathways in each environment using both the IPR and $C_{l1}$ as a measure of delocalisation.

\begin{table}[hbt!]
    \small
    \centering
    \begin{tabular}{lcccc}
    \hline
        \textrm{LH2 environment} & $\sqrt{\overline{C_{l1}^{\mathrm{B}800}}\ \overline{C_{l1}^{\mathrm{B}850}}}V_{\mathrm{B}800, \alpha_{2}}$    & $\sqrt{\overline{IPR^{\mathrm{B}800}}\ \overline{IPR^{\mathrm{B}850}}}V_{\mathrm{B}800, \alpha_{2}}$    & $\sqrt{\overline{C_{l1}^{\mathrm{B}800}}\ \overline{C_{l1}^{\mathrm{B}850}}}$  & $\sqrt{\overline{IPR^{\mathrm{B}800}}\ \overline{IPR^{\mathrm{B}850}}}$ \\
    \hline
    Membrane POPC   & 285     & 215     & 6.8    & 6    \\
    \hline
    Membrane DOPC   & 264     & 191     & 6.9    & 5     \\
    \hline
    Detergent       & 152     & 135     & 4.8     & 4      \\
    \hline
    \end{tabular}
\caption{Figure of merit that captures the electronic properties of the LH2 through the electronic coupling between the B800 and B850 ring and the delocalisation of excitons within each ring. The figure of merit for the dominating exciton transfer pathways in each environment. The increase in the transfer rate from detergent to membrane correlates with the increase in the figure of merit, as expected since changes in transfer rate between each environment are strongly dependent on changes in the electronic properties of the LH2. }
\label{tab:5}
\end{table}

The increase in the figure of merit correlates with the increase in exciton transfer rate of the dominant exciton pairs from detergent to POPC membrane. It allows us to relate the change in transfer rate directly to the B800 to B850 interchromophore electronic couplings and delocalisation. For POPC membrane, although $V_{B800, \alpha_{2}}$ is largest, the geometric average IPR of the B800 and B850 exciton pairs suggests that an increased delocalisation of the B850 and B800 excitons in POPC membrane also contributes to an increased exciton transfer rate. 

In the case of the non-dominant excitons, although the exciton coupling is stronger in POPC than DOPC membrane, the transfer rate is faster in DOPC. There is instead a correlation between the exciton transfer rate and the spectral overlap between the exciton line shape functions. The spectral overlap is dependent on the energy gap between the excitons and the width of the line shape which is determined by the real part of $g_{\alpha}(t)$. $g_{\alpha}(t)$ is given by $g_{i}(t)$ scaled by $1/\mathrm{IPR}$. In detergent, the B800 exciton is highly localised with an IPR of 3 resulting in a broader line shape. Additionally, the energy gap between the donor and acceptor exciton is smallest in detergent, resulting in a greater spectral overlap. Thus we see that for the non-dominant pathways, the spectral overlap is the dominating factor in determining the exciton transfer rate.

These results suggest that although there is an interplay between the exciton coupling strength and spectral overlap when determining the exciton transfer rate, the dominating transfer pathways in B800 to B850 transfer depend strongly on exciton coupling strength alone. Thus the real interplay is between the B800 to B850 interring coupling, and the delocalisation of excitons in each ring, two properties that can be traced back to the electronic properties of the LH2. Thus, changes in the electronic properties from detergent to membrane environments alter exciton energy transfer pathways, impacting the overall B800 to B850 transfer rate.


\section*{Discussion}

So far, knowledge of the structure and function of the LH2 complex has been gained mostly through investigating complexes isolated from their native environment in the photosynthetic membrane. Recent experimental studies have found that energy transfer within the complex is faster in a membrane environment that mimics the bacterial membrane \cite{ogren2018impact}. Using two levels of theory, namely, GFT and HEOM to estimate B800 to B850 energy transfer rates, we have been able to show how faster energy transfer in membrane-embedded LH2 can be linked to changes in the electronic properties of the complex. In agreement with previous theoretical studies, we have identified that the dominating pathway an excitation takes from the B800 to the B850 ring is via the dark B850* states, and find this to be the case in both membrane and detergent environments \cite{novoderezhkin2003intra}. We have shown how faster energy transfer in the membrane is the result of the increased delocalisation and stronger coupling between the excitons involved in the dominating pathways. Signatures of stronger electronic coupling in  the B850 ring are additionally present in both experimental and theoretical linear spectra which show a red shift in the B850 absorption, a change characteristic of stronger interchromophore electronic couplings \cite{ogren2018impact, freiberg2012comparative, agarwal2002nature}. Finally, we find a broader distribution of B800 to B850 energy transfer rates for an ensemble of 10,000 LH2’s and suggest that a biological trade-off may be present that allows the LH2 to achieve faster average energy transfer in membrane by having a broad spread of energy transfer rates.

We use both GFT and HEOM to determine the average B800 to B850 energy transfer rate in each environment and find a qualtiative agreement between the two approaches indicating a faster transfer rate in membrane LH2, in agreement with experimental pump-probe measurements \cite{ogren2018impact}. The authors suggest that lipid bilayer properties such as the lateral pressure profile of the membrane or a hydrophobic (mis)match may be the microscopic origin of the increased energy transfer rate in membrane. The membrane lipid bilayer provides stability to the LH2 complex through lateral pressure, which is altered when in detergent or in varying membrane lipid compositions \cite{andersen2007bilayer, ding2015effects}. To assess the importance of lipid-protein interactions on the energy transfer within the LH2, we computed the average B800 to B850 transfer rate for LH2 embedded in two different lipids, POPC and DOPC. Both GFT and HEOM rates predict slower transfer in DOPC compared to POPC indicating that changes in the lipid composition can result in changes in energy transfer. Understanding how the energy transfer within the LH2 changes as a function of the lipid properties could reveal how the complex achieves faster rates of energy transfer in a lipid environment. Although energy transfer is expected to be slowest in detergent as predicted by the HEOM derived rate, GFT predicts transfer to be slowest in DOPC. This discrepancy highlights the importance of  non-perturbative frameworks in order to resolve and understand the differences of energy transfer kinetics and to rationalise experimental observations. We note that the estimated transfer times with HEOM  are in general longer than with GFT. These quantitative differences result in part from the fact that in the HEOM  approach we map the  B800 to B850 transfer process onto the kinetics on a two state system. Rigourous approaches to extracting  more accurate transfer rates from a non-perturbative framework such as HEOM is an open problem which goes beyond the scope of the current manuscript and will be presented elsewhere.

An advantage of GFT is that the underlying excitonic properties can be studied to pinpoint changes that could result in faster B800 to B850 energy transfer. The dominating transfer pathways in the LH2 indicate that faster exciton transfer in the membrane can be linked to stronger excitonic couplings, a direct result of both stronger interchromophore electronic couplings and increased delocalisation of excitations. The membrane lipid bilayer provides stability to the LH2 complex through lateral pressure, which is altered when in detergent. Differences in lipid-protein and detergent-protein interactions could alter electronic couplings via perturbations in the geometry of the chromophores, as the protein scaffold has control over the position and orientation of the chromophores \cite{}.  Methods more sophisticated than the dipole-dipole approximation are used to determine all the electronic couplings in the membrane LH2 model that account for screening due to the lipid environment, such that a closer packing of B850 chromophores may not be the sole reason for stronger electronic couplings \cite{cupellini2016ab}. Accompanying stronger interchromophore electronic couplings, is the increased delocalisation of the excitons dominating B800 to B850 energy transfer in membrane. Two-dimensional spectroscopy measurements have found quantum beating in the fluorescence signals of the LH2 from \textit{R. acidophilus}, a signature of quantum coherent dynamics \cite{harel2012quantum}. Quantum beating signals arise as a result of constructive and destructive interference between different donor to single acceptor pathways over time. Such interference becomes possible when excitons are highly delocalised such that many relaxation pathways are available. Theoretical studies of model exciton systems suggest that interference is important to achieve high energy trapping efficiency suggesting coherent dynamics is important for. Interference becomes possible when excitons are highly delocalised such that many relaxation pathways are available. Thus the increased delocalisation of excitons in membrane LH2 could impact the coherent dynamics within the complex.

The dark B850* states seem to be an important energy acceptor for B800 to B850 energy transfer within the LH2 in both detergent and membrane environments. Previous theoretical studies have found that B800 to B850* energy transfer occurs faster (600 - 800fs) than transfer to lower energy bright B850 exciton states (1 ps) \cite{novoderezhkin2003intra} . Energy transfer pathways in the LH2 have been probed experimentally using two-dimensional spectroscopy, but It is difficult to detect a B800 to B850* signal since the third order nonlinear response measured is proportional to the fourth power of the transition dipole moment, which in the case of B850* is negligible \cite{fidler2013probing}. Fidler \textit{et al.} suggests that the excitons involved in this energy transfer pathway may have parallel transition dipole moments which would also prevent their detection. Despite its elusiveness, the presence of a fast B800 to B850* energy transfer pathway could explain the additional fast decay channel found when exciting LH2 at the blue end of the B800 band in hole burning experiments \cite{wu1996femtosecond, de1994inter, monshouwer1995low}. A similar pathway has been suggested in the LH3 complex from \textit{R. acidophilus} strain 7050, a low light variant of \textit{R. acidophilus} \cite{tong2020comparison}. The LH3 has a similar nonameric structure to the LH2 but with the B850 band blue shifted to 820 nm, suggesting that this mechanism may be shared across different variants of the complex. Studies on artificial light harvesting systems have shown that transfer of excitation energy from bright to dark states may be used to prevent re-emission since the dark state cannot optically decay, thus increasing the efficiency of energy transfer in the system \cite{creatore2013efficient, higgins2017quantum, zhang2016dark}. The dark B850* states may play a similar role in trapping absorbed solar energy by quickly moving excitation energy out of the B800 ring, where it would otherwise relax to low energy B800 states that have a greater transition dipole strength.

By calculating the B800 to B850 energy transfer rate for 10,000 realisations of static disorder, we also resolve the heterogeneity across an ensemble of LH2 complexes and can study the form of their distribution. We find a broader distribution of energy transfer rates in membrane embedded LH2 suggesting that the energy transfer mechanism has a lower level of accuracy in the native environment due to the greater standard deviation of the transfer rates compared to detergent. A concept used to understand the relationship between different traits in a biological system is a trade-off, which can be identified by a negative relationship between two traits. The distribution of the transfer rate in membrane and detergent LH2 suggests that the complex sacrifices accuracy in the transfer rate for speed. The traits underlying a speed-accuracy trade-off would likely be related to properties of the LH2 that change from detergent to membrane. However identifying a trade-off would require more thorough investigation and laboratory evolution experiments.

In summary, we have shown that increased energy transfer efficiency within membrane-embedded LH2 can be traced back to altered energy transfer pathways and enhanced quantum delocalisation of excitations within the complex. Further work towards understanding the biological interactions underlying such enhancements will not only provide a deeper understanding of the function of the LH2 but will also lead towards improved theoretical tools to study similar photosynthetic complexes. Currently there is a lack of comprehensive electronic and spectral density parameters available for membrane-embedded LH2 making such an investigation challenging. Additionally, to study the LH2 in its biological environment having a non-perturbative framework that can yield excitation transfer rates is essential. The work presented here is a first step towards addressing these challenges and uncovering the role that the biological environment plays in the efficiency of these light harvesting complexes.

\section*{Author Contributions}

AOC designed and supervised the research. CK and HÓG carried out the simulations. LC and BM provided quantum chemical insight. CK wrote the first draft. All authors analysed the data, discussed the results, and contributed to the final version of the manuscript. 

\section*{Acknowledgments}
We thank Gabriela Schlau-Cohen and Charlie Nation for insightful discussions. We gratefully acknowledge financial support from the Engineering and Physical Sciences Research Council (EPSRC UK) Grant EP/T517793/1 and from the Gordon and Betty Moore Foundation Grant 8820.  The authors acknowledge use of the UCL Myriad High Performance Computing Facility.

\bibliography{bibliography}

\section*{Supplementary Material}

\subsection*{Modified Redfield theory}

The energy transfer rates computed with GFT use lineshape functions that are derived perturbatively as introduced in the \nameref{Sec:GFTTheory} section. The exciton lifetime $\tau_{\alpha}$ that enters the lineshape functions is given by
\begin{equation}
    \tau_{\alpha}=\left(\frac{1}{2}\sum_{\alpha\neq\beta}k_{\alpha\beta}^{\mathrm{MR}}\right)^{-1},
\end{equation}
where $k_{\alpha\beta}^{\mathrm{MR}}$ is the energy transfer rate from a donor exciton $\alpha$ to all possible acceptor excitons $\beta$, given that they are localised on the same ring as the donor exciton. Since the interchromophore electronic couplings within each ring are strong, modified Redfield theory is used to obtain the intra-ring exciton transfer rates. By assuming that the electronic states of each ring are weakly coupled to their environment, $H_{\mathrm{SB}}$ is treated as a perturbation on the dynamics within each ring. 

The modified Redfield energy transfer rate between two excitons in the same ring is given by \cite{yang2002influence}
\begin{equation}
\begin{aligned}
    k_{\alpha\beta}^{\mathrm{MR}}=2\mathrm{Re}\int_{0}^{\infty} dt\ & e^{-i\omega_{\alpha\beta}t} e^{-i(\lambda_{\alpha\alpha,\alpha\alpha}+\lambda_{\beta\beta, \beta\beta})t} e^{-g_{\alpha}(t)-g_{\beta}(t)} e^{2g_{\beta\beta, \alpha\alpha}+2i\lambda_{\beta\beta, \alpha\alpha}} \\
    &\times \left[\ddot{g}_{\beta\alpha,\beta\alpha}(t)-(\dot{g}_{\beta\alpha,\beta\beta}(t)-\dot{g}_{\beta\alpha,\alpha\alpha}(t)+2i\lambda_{\beta\alpha,\beta\beta})^2\right],
\end{aligned}
\end{equation}
where the terms have been defined in the main text (see \nameref{Sec:GFTTheory} section).

\subsection*{Propagation of the dipole operator}
Numerical computation of the absorption and fluorescence expressions given by Eqs.\ (\ref{eq:absorption_spectra_expression}) and (\ref{eq:fluorescence_spectra_expression}) using HEOM theory is achieved by rewriting the auto-correlation as
\begin{equation}
    \braket{\hat{\mu}_p(t)\hat{\mu}_p(t)}_\rho = \text{Tr}(\hat{\mu}_p e^{\mathcal{L}t}[\hat{\mu}_p\hat{\rho}]),
\end{equation}
where $\mathcal{L}$ is the HEOM generator of dynamics and $\hat{\rho}$ is the reduced system density matrix. The half-sided Fourier transform is then formally calculated to give
\begin{equation}
    \int_0^\infty \text{d}t \hat{\mu}_p e^{\mathcal{L}t}[\hat{\mu}_p\hat{\rho}] e^{i\omega t} = -\hat{\mu}_p \frac{1}{\mathcal{L} + i\omega}[\hat{\mu}_p\hat{\rho}].
\end{equation}
We numerically determine $\hat{x}_{p,\omega} = \frac{1}{\mathcal{L} + i\omega}[\hat{\mu}_p\hat{\rho}]$ by solving the linear system $(\mathcal{L} + i\omega)[\hat{x}_{p,\omega}] = \hat{\mu}_p\hat{\rho}$ using the BiCGSTAB Krylov subspace method \cite{vanderVorst1992Mar}. This method of numerically computing spectra is more efficient than numerically Fourier transforming the dynamics as a result of the sparsity of the matrix representation for $\mathcal{L}$. 

\subsection*{Finding the thermal state}
The fluorescence spectra in Eq.\ \eqref{eq:fluorescence_spectra_expression} is computed by performing a trace with respect to the thermal state $\hat{\rho}_{th}$, which satisfies the property $\mathcal{L}\hat{\rho}_{th} = 0$. In order to determine the thermal state we solve this linear system using the BiCGSTAB method \cite{vanderVorst1992Mar} which is supplied with an initial guess given by the Boltzmann state $e^{-\beta\hat{H}} / \text{Tr}(e^{-\beta\hat{H}})$. Doing so guarantees that the solver will not yield the trivial zero matrix solution, which of course does not represent a physical state. 

\subsection*{B800 to B850 transfer rate distributions}

\end{document}